\documentclass[longauth]{aa}  

\usepackage{graphicx}
\usepackage{txfonts}
\usepackage{dcolumn}
\usepackage{braket}
\usepackage{tabularx,rotating,multirow}
\usepackage{float}
\usepackage[caption=false]{subfig}
\usepackage{hyperref}
\usepackage{amsfonts}

\makeatletter
\renewcommand*\aa@pageof{, page \thepage{} of \pageref*{LastPage}}
\makeatother

\begin{document}

   \title{\emph{Euclid}: The reduced shear approximation and magnification bias for Stage IV cosmic shear experiments\thanks{This paper is published on behalf of the Euclid Consortium.}}


\author{A.C.~Deshpande$^{1}$\thanks{\email{anurag.deshpande.18@ucl.ac.uk}}, T.D.~Kitching$^{1}$, V.F.~Cardone$^{2,3}$, P.L.~Taylor$^{1,4}$, S.~Casas$^{5}$, S.~Camera$^{6,7,8}$, C.~Carbone$^{9,10}$, M.~Kilbinger$^{5,11}$, V.~Pettorino$^{5}$, Z.~Sakr$^{12,13}$, D.~Sapone$^{14}$, I.~Tutusaus$^{13,15,16}$, N.~Auricchio$^{17}$, C.~Bodendorf$^{18}$, D.~Bonino$^{8}$, M.~Brescia$^{19}$, V.~Capobianco$^{8}$, J.~Carretero$^{20}$, M.~Castellano$^{3}$, S.~Cavuoti$^{19,21,22}$, R.~Cledassou$^{23}$, G.~Congedo$^{24}$, L.~Conversi$^{25}$, L.~Corcione$^{8}$, M.~Cropper$^{1}$, F.~Dubath$^{26}$, S.~Dusini$^{27}$, G.~Fabbian$^{28}$, E.~Franceschi$^{17}$, M.~Fumana$^{10}$, B.~Garilli$^{10}$, F.~Grupp$^{18}$, H.~Hoekstra$^{29}$, F.~Hormuth$^{30}$, H.~Israel$^{31}$, K.~Jahnke$^{32}$, S.~Kermiche$^{33}$, B.~Kubik$^{34}$, M.~Kunz$^{35}$, F.~Lacasa$^{36}$, S.~Ligori$^{8}$, P.B.~Lilje$^{37}$, I.~Lloro$^{15,16}$, E.~Maiorano$^{17}$, O.~Marggraf$^{38}$, R.~Massey$^{39}$, S.~Mei$^{40}$, M.~Meneghetti$^{17}$, G.~Meylan$^{41}$, L.~Moscardini$^{17,42,43}$, C.~Padilla$^{20}$, S.~Paltani$^{26}$, F.~Pasian$^{44}$, S.~Pires$^{5}$, G.~Polenta$^{45}$, M.~Poncet$^{23}$, F.~Raison$^{18}$, J.~Rhodes$^{4}$, M.~Roncarelli$^{17,42}$, R.~Saglia$^{18}$, P.~Schneider$^{38}$, A.~Secroun$^{33}$, S.~Serrano$^{16,46}$, G.~Sirri$^{47}$, J.L.~Starck$^{5}$, F.~Sureau$^{5}$, A.N.~Taylor$^{24}$, I.~Tereno$^{48,49}$, R.~Toledo-Moreo$^{50}$, L.~Valenziano$^{17,47}$, Y.~Wang$^{51}$, J.~Zoubian$^{33}$}
   
\institute{$^{1}$ Mullard Space Science Laboratory, University College London, Holmbury St Mary, Dorking, Surrey RH5 6NT, UK\\
$^{2}$ I.N.F.N.-Sezione di Roma Piazzale Aldo Moro, 2 - c/o Dipartimento di Fisica, Edificio G. Marconi, I-00185 Roma, Italy\\
$^{3}$ INAF-Osservatorio Astronomico di Roma, Via Frascati 33, I-00078 Monteporzio Catone, Italy\\
$^{4}$ Jet Propulsion Laboratory, California Institute of Technology, 4800 Oak Grove Drive, Pasadena, CA, 91109, USA\\
$^{5}$ AIM, CEA, CNRS, Universit\'{e} Paris-Saclay, Universit\'{e} Paris Diderot, Sorbonne Paris Cit\'{e}, F-91191 Gif-sur-Yvette, France\\
$^{6}$ INFN-Sezione di Torino, Via P. Giuria 1, I-10125 Torino, Italy\\
$^{7}$ Dipartimento di Fisica, Universit\'a degli Studi di Torino, Via P. Giuria 1, I-10125 Torino, Italy\\
$^{8}$ INAF-Osservatorio Astrofisico di Torino, Via Osservatorio 20, I-10025 Pino Torinese (TO), Italy\\
$^{9}$ INFN-Sezione di Milano, Via Celoria 16, I-20133 Milano, Italy\\
$^{10}$ INAF-IASF Milano, Via Alfonso Corti 12, I-20133 Milano, Italy\\
$^{11}$ Institut d'Astrophysique de Paris, 98bis Boulevard Arago, F-75014, Paris, France\\
$^{12}$ Universit\'e St Joseph; UR EGFEM, Faculty of Sciences, Beirut, Lebanon\\
$^{13}$ Institut de Recherche en Astrophysique et Plan\'etologie (IRAP), Universit\'e de Toulouse, CNRS, UPS, CNES, 14 Av. Edouard Belin, F-31400 Toulouse, France\\
$^{14}$ Departamento de F\'isica, FCFM, Universidad de Chile, Blanco Encalada 2008, Santiago, Chile\\
$^{15}$ Institute of Space Sciences (ICE, CSIC), Campus UAB, Carrer de Can Magrans, s/n, 08193 Barcelona, Spain\\
$^{16}$ Institut d’Estudis Espacials de Catalunya (IEEC), 08034 Barcelona, Spain\\
$^{17}$ INAF-Osservatorio di Astrofisica e Scienza dello Spazio di Bologna, Via Piero Gobetti 93/3, I-40129 Bologna, Italy\\
$^{18}$ Max Planck Institute for Extraterrestrial Physics, Giessenbachstr. 1, D-85748 Garching, Germany\\
$^{19}$ INAF-Osservatorio Astronomico di Capodimonte, Via Moiariello 16, I-80131 Napoli, Italy\\
$^{20}$ Institut de F\'isica d’Altes Energies IFAE, 08193 Bellaterra, Barcelona, Spain\\
$^{21}$ Department of Physics "E. Pancini", University Federico II, Via Cinthia 6, I-80126, Napoli, Italy\\
$^{22}$ INFN section of Naples, Via Cinthia 6, I-80126, Napoli, Italy\\
$^{23}$ Centre National d'Etudes Spatiales, Toulouse, France\\
$^{24}$ Institute for Astronomy, University of Edinburgh, Royal Observatory, Blackford Hill, Edinburgh EH9 3HJ, UK\\
$^{25}$ ESAC/ESA, Camino Bajo del Castillo, s/n., Urb. Villafranca del Castillo, 28692 Villanueva de la Ca\~nada, Madrid, Spain\\
$^{26}$ Department of Astronomy, University of Geneva, ch. d'\'Ecogia 16, CH-1290 Versoix, Switzerland\\
$^{27}$ INFN-Padova, Via Marzolo 8, I-35131 Padova, Italy\\
$^{28}$ Department of Physics \& Astronomy, University of Sussex, Brighton BN1 9QH, UK\\
$^{29}$ Leiden Observatory, Leiden University, Niels Bohrweg 2, 2333 CA Leiden, The Netherlands\\
$^{30}$ von Hoerner \& Sulger GmbH, Schlo{\ss}Platz 8, D-68723 Schwetzingen, Germany\\
$^{31}$ Universit\"ats-Sternwarte M\"unchen, Fakult\"at f\"ur Physik, Ludwig-Maximilians-Universit\"at M\"unchen, Scheinerstrasse 1, 81679 M\"unchen, Germany\\
$^{32}$ Max-Planck-Institut f\"ur Astronomie, K\"onigstuhl 17, D-69117 Heidelberg, Germany\\
$^{33}$ Aix-Marseille Univ, CNRS/IN2P3, CPPM, Marseille, France\\
$^{34}$ Univ Lyon, Univ Claude Bernard Lyon 1, CNRS/IN2P3, IP2I Lyon, UMR 5822, F-69622, Villeurbanne, France\\
$^{35}$ Universit\'e de Gen\`eve, D\'epartement de Physique Th\'eorique and Centre for Astroparticle Physics, 24 quai Ernest-Ansermet, CH-1211 Gen\`eve 4, Switzerland\\
$^{36}$ Institut d'Astrophysique Spatiale (IAS), B\^atiment 121, F-91405 Orsay, Universit\'e Paris-Sud 11 and CNRS, UMR 8617, France\\
$^{37}$ Institute of Theoretical Astrophysics, University of Oslo, P.O. Box 1029 Blindern, N-0315 Oslo, Norway\\
$^{38}$ Argelander-Institut f\"ur Astronomie, Universit\"at Bonn, Auf dem H\"ugel 71, 53121 Bonn, Germany\\
$^{39}$ Institute for Computational Cosmology, Department of Physics, Durham University, South Road, Durham, DH1 3LE, UK\\
$^{40}$ Universit\'{e} de Paris, F-75013, Paris, France, LERMA, Observatoire de Paris, PSL Research University, CNRS, Sorbonne Universit\'e, F-75014 Paris, France\\
$^{41}$ Observatoire de Sauverny, Ecole Polytechnique F\'ed\'erale de Lau- sanne, CH-1290 Versoix, Switzerland\\
$^{42}$ Dipartimento di Fisica e Astronomia, Universit\'a di Bologna, Via Gobetti 93/2, I-40129 Bologna, Italy\\
$^{43}$ INFN-Bologna, Via Irnerio 46, I-40126 Bologna, Italy\\
$^{44}$ INAF-Osservatorio Astronomico di Trieste, Via G. B. Tiepolo 11, I-34131 Trieste, Italy\\
$^{45}$ Space Science Data Center, Italian Space Agency, via del Politecnico snc, 00133 Roma, Italy\\
$^{46}$ Institute of Space Sciences (IEEC-CSIC), c/Can Magrans s/n, 08193 Cerdanyola del Vall\'es, Barcelona, Spain\\
$^{47}$ INFN-Sezione di Bologna, Viale Berti Pichat 6/2, I-40127 Bologna, Italy\\
$^{48}$ Instituto de Astrof\'isica e Ci\^encias do Espa\c{c}o, Faculdade de Ci\^encias, Universidade de Lisboa, Tapada da Ajuda, PT-1349-018 Lisboa, Portugal\\
$^{49}$ Departamento de F\'isica, Faculdade de Ci\^encias, Universidade de Lisboa, Edif\'icio C8, Campo Grande, PT1749-016 Lisboa, Portugal\\
$^{50}$ Universidad Polit\'ecnica de Cartagena, Departamento de Electr\'onica y Tecnolog\'ia de Computadoras, 30202 Cartagena, Spain\\
$^{51}$ Infrared Processing and Analysis Center, California Institute of Technology, Pasadena, CA 91125, USA\\
}

   \date{Received December 16, 2019/ Accepted March 04, 2020}

 
  \abstract
   {Stage IV weak lensing experiments will offer more than an order of magnitude leap in precision. We must therefore ensure that our analyses remain accurate in this new era. Accordingly, previously ignored systematic effects must be addressed.}
   {In this work, we evaluate the impact of the reduced shear approximation and magnification bias on information obtained from the angular power spectrum. To first-order, the statistics of reduced shear, a combination of shear and convergence, are taken to be equal to those of shear. However, this approximation can induce a bias in the cosmological parameters that can no longer be neglected. A separate bias arises from the statistics of shear being altered by the preferential selection of galaxies and the dilution of their surface densities in high-magnification regions.}
   {The corrections for these systematic effects take similar forms, allowing them to be treated together. We calculated the impact of neglecting these effects on the cosmological parameters that would be determined from \emph{Euclid}, using cosmic shear tomography. To do so, we employed the Fisher matrix formalism, and included the impact of the super-sample covariance. We also demonstrate how the reduced shear correction can be calculated using a lognormal field forward modelling approach.}
   {These effects cause significant biases in $\Omega_{\rm m}$, $\sigma_{8}$, $n_{\rm s}$, $\Omega_{\rm DE}$, $w_0$, and $w_a$ of $-0.53\sigma$, $0.43\sigma$, $-0.34\sigma$, $1.36\sigma$, $-0.68\sigma$, and $1.21\sigma$, respectively. We then show that these lensing biases interact with another systematic effect: the intrinsic alignment of galaxies. Accordingly, we have developed the formalism for an intrinsic alignment-enhanced lensing bias correction. Applying this to \emph{Euclid}, we find that the additional terms introduced by this correction are sub-dominant.}
   {}

   \keywords{Cosmology: observations --
                Gravitational lensing: weak --
                Methods: analytical
               }
    
  \authorrunning{A.C.~Deshpande et al.} 
  \maketitle
%

\section{\label{sec:level1}Introduction}

The constituent parts of the Lambda cold dark matter ($\Lambda$CDM) model, and its extensions, are not all fully understood. In the current framework, there is no definitive explanation for the physical natures of dark matter and dark energy. Today, there are a variety of techniques available to better constrain our knowledge of the $\Lambda$CDM cosmological parameters. Cosmic shear, which is the distortion in the observed shapes of distant galaxies due to weak gravitational lensing by the large-scale structure of the Universe (LSS), is one such cosmological probe. By measuring this distortion over large samples of galaxies, the LSS can be explored. Given that the LSS depends on density fluctuations and the geometry of the Universe, this measurement allows for the constraining of cosmological parameters. In particular, it is a powerful tool to study dark energy \citep{DETFrep}. A three-dimensional, redshift-dependent, picture can be obtained using a technique known as tomography. In this technique, the observed galaxies are divided into different tomographic bins; each covering a different redshift range.

Since its debut at the turn of the millennium \citep{Bacon2000,Kaiser2000,VanWaerbeke2000,Wittman2000, Rhodes2000}, studies of cosmic shear have evolved to the point where multiple independent surveys have carried out precision cosmology \citep{DESpap, cfhtmain, kids450}. Now, with the impending arrival of Stage IV \citep{DETFrep} dark energy experiments, such as \emph{Euclid}\footnote{\url{https://www.euclid-ec.org/}} \citep{EuclidRB}, \emph{WFIRST}\footnote{\url{https://www.nasa.gov/wfirst}} \citep{WFIRSTpap}, and LSST\footnote{\url{https://www.lsst.org/}} \citep{LSSTpap}, we are poised for a leap in precision. For example, even a pessimistic analysis of \emph{Euclid} weak lensing data is projected to increase precision by a factor of $\sim$25 over current surveys \citep{SellentinStarck19}.

To ensure that the accuracy of the analysis keeps up with the increasing precision of the measurements, the impact of previously neglected physical effects must be evaluated. In cosmic shear a wide range of scales are probed, so the non-linear matter power spectrum must be precisely modelled. This can be accomplished through model fitting to N-body simulations \citep{Smith03,Takahashi12}. A robust understanding of how baryonic physics affects the matter power spectrum is also necessary \citep{RuddBaryon, Owlsbisup}. Furthermore, spurious signals arising from intrinsic alignments (IAs) \citep{IA1,IA2,IA3} in observed galaxy shapes need to be taken into account.

Additionally, assumptions in the theoretical formalism must also be relaxed. The effects of several such extensions on a \emph{Euclid}-like experiment have been investigated. These include: the impacts of relaxing the Limber, Hankel transform and flat-sky approximations \citep{limitsofshear17}, of using unequal-time correlators \citep{UTCpap}, and of making the spatially-flat Universe approximation \citep{Spatflatpap}. 

The formalism to correct for the effect of measuring reduced shear, rather than shear itself, is known \citep{Shapiro09,KrauseHirataRS}. However, its impact on impending surveys has not yet been quantified. The correction to the two-point cosmic shear statistic for magnification bias is also known. While the impact of this on Stage IV experiments has been quantified in \cite{LSSTMagBias}, the approach taken here risks underestimating the bias for surveys covering the redshift range of \emph{Euclid}. Rather than assuming that the magnification bias at the survey's mean redshift is representative of the bias at all covered redshifts, a tomographic approach is required. Magnification bias also affects measurements of galaxy clustering; which is the other of \emph{Euclid}'s major probes. \cite{thielmag, Lormag, DuncMag} study the impact of magnification on the clustering sample, and as such complement this work in forming a holistic picture of the effect on \emph{Euclid}.

Conveniently, the magnification bias correction takes a mathematically similar form to that of reduced shear; meaning these corrections can be treated together \citep{RSMBcombpap}. Within this work, we calculate the bias on the predicted cosmological parameters obtained from \emph{Euclid}, when these two effects are neglected. We further extend the existing correction formalism to include the impact of IAs, and recompute the bias for this case. 

In Sect. \ref{sec:2}, we establish the theoretical formalism. We begin by summarising the standard, first-order, cosmic shear power spectrum calculation. We then review the basic reduced shear correction formalism of \cite{Shapiro09}. Following this, the correction for magnification bias is explained. Next, the theory used to account for the IAs is examined. We then combine the discussed schemes, in order to create a description of an IA-enhanced lensing bias correction to the cosmic shear power spectrum. We also explain how we quantify the uncertainties and biases induced in the measured cosmological parameters. 

In Sect. \ref{sec:3}, we describe how we calculated the impact of the aforementioned corrections for \emph{Euclid}. Our modelling assumptions and choice of fiducial cosmology are stated, and computational specifics are given. 

Finally, in Sect. \ref{sec:4}, our results are presented, and their implications for \emph{Euclid} are discussed. The biases and change in confidence contours of cosmological parameters, resulting from the basic reduced shear and magnification bias corrections, are presented. We also present the biases from the IA-enhanced lensing bias correction. 

\section{\label{sec:2}Theoretical formalism}

Here, we first review the standard cosmic shear calculation. We then explain the corrections required to account for the reduced shear approximation, and for magnification bias. We further consider the effects of IAs, and construct an IA-enhanced lensing bias correction. The formalism for accounting for the shot noise is then stated. Our chosen framework for predicting uncertainties and biases is also detailed.

\subsection{\label{sec:21}The standard cosmic shear calculation}

When a distant galaxy is weakly lensed, the change in its observed ellipticity is proportional to the reduced shear, \textit{g}:
\begin{equation}
    \label{eq:redshear}
    g^\alpha(\boldsymbol{\theta})= \frac{\gamma^\alpha(\boldsymbol{\theta})}{1-\kappa(\boldsymbol{\theta})},
\end{equation}
where $\boldsymbol{\theta}$ is the galaxy's position on the sky, $\gamma$ is the shear, which is an anisotropic stretching that turns circular distributions of light elliptical, and $\kappa$ is the convergence, which is an isotropic change in the size of the image. The superscript, $\alpha$, encodes the fact that the spin-2 shear has two components. Since $|\gamma|$, $|\kappa|$ $\ll$ 1 for individual galaxies in weak lensing, Eq. (\ref{eq:redshear}) is typically approximated to first-order as $g^\alpha(\boldsymbol{\theta}) \approx \gamma^\alpha(\boldsymbol{\theta})$. This is known as the reduced shear approximation.

The convergence of a source being weakly lensed by the LSS, in a tomographic redshift bin \textit{i}, is given by: 
\begin{equation}
    \label{eq:convergence}
    \kappa_i(\boldsymbol{\theta})=\int_{0}^{\chi_{\rm lim}} {\rm d}\chi\:\delta[d_{\rm A}(\chi)\boldsymbol{\theta},\, \chi]\:W_i(\chi).
\end{equation}
It is the projection of the density contrast of the Universe, $\delta$, over the comoving distance, $\chi$, along the line-of-sight, to the limiting comoving distance of the observed sample of sources, $\chi_{\rm lim}$. The function \textit{$d_{\rm A}(\chi)$} accounts for the curvature of the Universe, $K$, depending on whether it is flat, open, or closed:
\begin{equation}
    \label{eq:dA}
    d_{\rm A}(\chi) = \begin{cases}
    |K|^{-1/2}\sin(|K|^{-1/2}\chi) & \text{\small{$K>0$ (Closed)}}\\
    \chi & \text{\small{$K=0$ (Flat)}}\\
    |K|^{-1/2}\sinh(|K|^{-1/2}\chi) & \text{\small{$K<0$ (Open)},}
  \end{cases}
\end{equation}
and \textit{W$_{i}(\chi)$} is the lensing kernel for sources in bin \textit{i}, with the definition
\begin{equation}
    \label{eq:Wi}
    W_i(\chi) = \frac{3}{2}\Omega_{\rm m}\frac{H_0^2}{c^2}\frac{d_{\rm A}(\chi)}{a(\chi)}\int_{\chi}^{\chi_{\rm lim}}{\rm d}\chi'\:n_i(\chi')\:\frac{d_{\rm A}(\chi'-\chi)}{d_{\rm A}(\chi')}.
\end{equation}
Here, $\Omega_{\rm m}$ is the dimensionless present-day matter density parameter of the Universe, $H_0$ is the Hubble constant, $c$ is the speed of light in a vacuum, $a(\chi)$ is the scale factor of the Universe, and $n_i(\chi)$ is the probability distribution of galaxies within bin $i$.

Meanwhile, the two shear components, for a bin $i$, when  caused by a lensing mass distribution, can be related to the convergence in a straightforward manner in frequency space:
\begin{equation}
    \label{eq:fourier}
    \widetilde{\gamma}_i^\alpha(\boldsymbol{\ell})=\frac{1}{\ell(\ell+1)}\sqrt{\frac{(\ell+2)!}{(\ell-2)!}}\:T^\alpha(\boldsymbol{\ell})\,\widetilde{\kappa}_i(\boldsymbol{\ell}),
\end{equation}
where $\boldsymbol{\ell}$ is the spherical harmonic conjugate of $\boldsymbol{\theta}$. Here, the small-angle limit is used. However, we do not apply the `prefactor unity' approximation \citep{limitsofshear17}, in which the factor of $1/\ell(\ell+1)\sqrt{(\ell+2)!/(\ell-2)!}$ is taken to be one, despite the fact that the impact of making the approximation is negligible for a \emph{Euclid}-like survey \citep{Kilbingerellfactors}. This is done to allow consistent comparison with the spherical-sky reduced shear and magnification bias corrections. The trigonometric weighting functions, $T^\alpha(\boldsymbol{\ell})$, of the two shear components are defined as:
\begin{align}
    \label{eq:Trigfunc1}
    T^1(\boldsymbol{\ell}) &= \cos(2\phi_\ell),\\
    \label{eq:Trigfunc2}
    T^2(\boldsymbol{\ell}) &= \sin(2\phi_\ell),
\end{align}
with $\phi_l$ being the angular component of vector $\boldsymbol{\ell}$ which has magnitude $\ell$. Then, for an arbitrary shear field (e.g. one estimated from data), the two shear components can be linearly combined to be represented as a curl-free $E$-mode, and a divergence-free $B$-mode:
\begin{align}
    \label{eq:Emode}
    \widetilde{E}_i(\boldsymbol{\ell})&=\sum_\alpha T^\alpha\:\widetilde{\gamma}_i^\alpha(\boldsymbol{\ell}),\\
    \label{eq:Bmode}
    \widetilde{B}_i(\boldsymbol{\ell})&=\sum_\alpha \sum_\beta \varepsilon^{\alpha\beta}\,T^\alpha(\boldsymbol{\ell})\:\widetilde{\gamma}_i^\beta(\boldsymbol{\ell}),
\end{align}
where $\varepsilon^{\alpha\beta}$ is the two-dimensional Levi-Civita symbol, so that $\varepsilon^{12}=-\varepsilon^{21}=1$ and $\varepsilon^{11}=\varepsilon^{22}=0$. The $B$-mode vanishes in the absence of any higher-order systematic effects. Therefore, we can then define the $E$-mode angular auto-correlation and cross-correlation spectra, $C_{\ell;ij}^{\gamma\gamma}$, as:
\begin{equation}
    \label{eq:powerspecdef}
    \braket{\widetilde{E}_i(\boldsymbol{\ell})\widetilde{E}_j(\boldsymbol{\ell'})} = (2\pi)^2\,\delta_{\rm D}^2(\boldsymbol{\ell}+\boldsymbol{\ell'})\,C_{\ell;ij}^{\gamma\gamma},
\end{equation}
where $\delta_{\rm D}^2$ is the two-dimensional Dirac delta function. From here, an expression is derived for $C_{\ell;ij}^{\gamma\gamma}$:
\begin{equation}
    \label{eq:Cl}
    C_{\ell;ij}^{\gamma\gamma} = \frac{(\ell+2)!}{(\ell-2)!}\frac{1}{(\ell+1/2)^4}\int_0^{\chi_{\rm lim}}{\rm d}\chi\frac{W_i(\chi)W_j(\chi)}{d^{\,2}_{\rm A}(\chi)}P_{\delta\delta}(k, \chi),
\end{equation}
where $P_{\delta\delta}(k, \chi)$ is the three-dimensional matter power spectrum. Obtaining Eq. (\ref{eq:Cl}) relies on making the Limber approximation, that is, assuming that only $\ell$-modes in the plane of the sky contribute to the lensing signal. Under the Limber approximation, $k=(\ell+1/2)/d_{\rm A}(\chi)$. In this equation, the factors of $(\ell+2)!/(\ell-2)!$ and  $1/(\ell+1/2)^4$ come once again from the fact that the prefactor unity approximation is not used. For a comprehensive review, see \cite{Kilbinger15}.

\subsection{\label{sec:22}The reduced shear correction}

We account for the effects of the reduced shear approximation by means of a second-order correction to Eq. (\ref{eq:Cl}) \citep{Shapiro09, KrauseHirataRS, DodelsonRS}. This can be done by taking the Taylor expansion of Eq. (\ref{eq:redshear}) around $\kappa=0$, and keeping terms up to second-order:
\begin{equation}
    \label{eq:gexpan}
    g^\alpha(\boldsymbol{\theta})=\gamma^\alpha(\boldsymbol{\theta})+(\gamma^\alpha\kappa)(\boldsymbol{\theta})+\mathcal{O}(\kappa^3).
\end{equation}
By substituting this expanded form of $g^\alpha$ for $\gamma^\alpha$ in Eq. (\ref{eq:Emode}) and then recomputing the $E$-mode ensemble average, we obtain the original result of Eq. (\ref{eq:powerspecdef}), plus a correction:
\begin{align}
    \label{eq:ecorr}
    \delta\braket{\widetilde{E}_i(\boldsymbol{\ell})\widetilde{E}_j(\boldsymbol{\ell'})} &= (2\pi)^2\,\delta_{\rm D}^2(\boldsymbol{\ell}+\boldsymbol{\ell'})\:\delta C^{\rm RS}_{\ell;ij} \nonumber\\ &=  \sum_\alpha \sum_\beta T^\alpha(\boldsymbol{\ell})T^\beta(\boldsymbol{\ell'})\braket{\widetilde{(\gamma^\alpha\kappa)}_i(\boldsymbol{\ell})\:\widetilde{\gamma}_j^\beta(\boldsymbol{\ell'})} \nonumber\\ &+ T^\alpha(\boldsymbol{\ell'})T^\beta(\boldsymbol{\ell})\braket{\widetilde{(\gamma^\alpha\kappa)}_j(\boldsymbol{\ell'})\:\widetilde{\gamma}_i^\beta(\boldsymbol{\ell})},
\end{align}
where $\delta C^{\rm RS}_{\ell;ij}$ are the resulting corrections to the angular auto and cross-correlation spectra. Applying the Limber approximation once more, we obtain an expression for these:
\begin{align}
    \label{eq:dCl}
    \delta C^{\rm RS}_{\ell;ij} &= \ell(\ell+1)\frac{(\ell+2)!}{(\ell-2)!}\frac{1}{(\ell+1/2)^6}\int_0^\infty\frac{{\rm d}^2\boldsymbol{\ell'}}{(2\pi)^2}\cos(2\phi_{\ell'}-2\phi_\ell) \nonumber\\
    &\times B_{ij}^{\kappa\kappa\kappa}(\boldsymbol{\ell}, \boldsymbol{\ell'}, -\boldsymbol{\ell}-\boldsymbol{\ell'}).
\end{align}
The factors of $\ell(\ell+1)(\ell+2)!/(\ell-2)!$ and $1/(\ell+1/2)^6$ arise from foregoing the three-point equivalent of the prefactor unity approximation. As in the case of Eq. (\ref{eq:fourier}), the product of these factors can be well approximated by one. However, we do not make this approximation for the sake of completeness, and as the additional factors do not add any significant computational expense. Here, $B_{ij}^{\kappa\kappa\kappa}$ is the two-redshift convergence bispectrum, which takes the following form:
\begin{align}
    \label{eq:bispecK}
    B_{ij}^{\kappa\kappa\kappa}(\boldsymbol{\ell_1}, \boldsymbol{\ell_2}, \boldsymbol{\ell_3}) &= B_{iij}^{\kappa\kappa\kappa}(\boldsymbol{\ell_1}, \boldsymbol{\ell_2}, \boldsymbol{\ell_3}) + B_{ijj}^{\kappa\kappa\kappa}(\boldsymbol{\ell_1}, \boldsymbol{\ell_2}, \boldsymbol{\ell_3})\nonumber\\ &=\int_0^{\chi_{\rm lim}}\frac{{\rm d}\chi}{d^{\,4}_{\rm A}(\chi)}W_i(\chi)W_j(\chi)[W_i(\chi)+W_j(\chi)]\nonumber\\
    &\times B_{\delta\delta\delta}(\boldsymbol{k_1},\boldsymbol{k_2},\boldsymbol{k_3},\chi),
\end{align}
where $B_{iij}^{\kappa\kappa\kappa}$ and $B_{ijj}^{\kappa\kappa\kappa}$ are the three-redshift bispectra, $k_x$ is the magnitude and $\phi_{\ell;x}$ is the angular component of $\boldsymbol{k_x}$ (for $x=1, 2, 3$). Under the Limber approximation, $k_x=(\ell_x+1/2)/d_{\rm A}(\chi)$. Here, we also approximate our photometric redshift bins to be infinitesimally narrow. In reality, because these bins would have a finite width, the product of lensing kernels in Eq. \ref{eq:bispecK} would be replaced by a single integral over the products of the contents of the integral in Eq. \ref{eq:Wi}. Accordingly, the values of the bispectrum would be slightly higher. However, given that \emph{Euclid} will have high quality photometric redshift measurement, we expect this difference to be negligible. Consequently, in our calculations we proceeded with the narrow-bin approximation, which allowed us to use the same lensing kernels as used in the power spectrum calculation. 

Analogous to the first-order power spectra being projections of the three-dimensional matter power spectrum, the two-dimensional convergence bispectra are a projection of the three-dimensional matter bispectrum, $B_{\delta\delta\delta}(\boldsymbol{k_1},\boldsymbol{k_2},\boldsymbol{k_3},\chi)$. The analytic form of the matter bispectrum is not well known. Instead, a semi-analytic approach starting with second-order perturbation theory \citep{Fryperturb}, and then fitting its result to N-body simulations, is employed. We used the fitting formula of \cite{Scoccimarro01}. Accordingly, the matter bispectrum can be written:
\begin{align}
    \label{eq:bispecFF}
    B_{\delta\delta\delta}(\boldsymbol{k_1},\boldsymbol{k_2},\boldsymbol{k_3},\chi) &= 2F_2^{\rm eff}\,(\boldsymbol{k_1},\boldsymbol{k_2})\,P_{\delta\delta}(k_1, \chi)P_{\delta\delta}(k_2, \chi) \nonumber\\
    &+ \text{cyc.} ,
\end{align}
where $F_2^{\rm eff}$ encapsulates the simulation fitting aspect, and is defined as:
\begin{align}
\label{eq:Feff}
        F_2^{\rm eff}(\boldsymbol{k_1},\boldsymbol{k_2}) &= \frac{5}{7}\,a(n_{\rm s},k_1)\,a(n_{\rm s},k_2) \nonumber\\
        &+ \frac{1}{2}\frac{\boldsymbol{k_1}\cdot\boldsymbol{k_2}}{k_1k_2}\,\bigg(\frac{k_1}{k_2}+\frac{k_2}{k_1}\bigg)\,b(n_{\rm s},k_1)\,b(n_{\rm s},k_2) \nonumber\\
        &+\frac{2}{7}\,\bigg(\frac{\boldsymbol{k_1}\cdot\boldsymbol{k_2}}{k_1k_2}\bigg)^2c(n_{\rm s},k_1)\,c(n_{\rm s},k_2),
\end{align}
where $n_{\rm s}$ is the scalar spectral index, which indicates the deviation of the primordial matter power spectrum from scale invariance ($n_{\rm s}=1$), and the functions $a,b$, and $c$ are fitting functions, defined in \cite{Scoccimarro01}. There are no additional correction terms of form $\widetilde{E}\widetilde{B}$ or $\widetilde{B}\widetilde{B}$, and it has been shown that higher-order terms are sub-dominant \citep{KrauseHirataRS}, so further terms in Eq. (\ref{eq:gexpan}) can be neglected for now.

\subsection{\label{sec:22.5}The magnification bias correction}

The observed overdensity of galaxies on the sky is affected by gravitational lensing in two competing ways \citep{MBorig}. Firstly, individual galaxies can be magnified (or demagnified), which results in their flux being increased (or decreased). At the flux limit of a survey, this can cause fainter sources (which in the absence of lensing would be excluded) to be included in the observed sample. Conversely, the density of galaxies in the patch of sky around this source appears reduced (or increased) due to the patch of sky being magnified (or demagnified) similarly to the source. Accordingly, the net effect of these depends on the slope of the intrinsic, unlensed, galaxy luminosity function, at the survey's flux limit. This net effect is known as magnification bias. Additionally, galaxies can also be pulled into a sample because their effective radius is increased as a consequence of magnification, such that they pass a resolution factor cut. In this work, we do not consider this effect as it is more important for ground-based surveys than space-based ones such as \emph{Euclid}.

In the case of weak lensing, where $|\kappa| \ll 1$, and assuming that fluctuations in the intrinsic galaxy overdensity are small on the scales of interest, the observed galaxy overdensity in tomographic bin $i$ is \citep{MBcorssource, MBorig}:
\begin{equation}
    \label{eq:galover}
    \delta^g_{{\rm obs}; i}(\boldsymbol{\theta}) = \delta^g_i(\boldsymbol{\theta}) + (5s_i-2)\kappa_i(\boldsymbol{\theta}),
\end{equation}
where $\delta^g_i(\boldsymbol{\theta})$ is the intrinsic, unlensed, galaxy overdensity in bin $i$, and $s_i$ is the slope of the cumulative galaxy number counts brighter than the survey's limiting magnitude, $m_{\rm lim}$, in redshift bin $i$. This slope is defined as:
\begin{equation}
    \label{eq:slope}
    s_i = \frac{\partial{\rm log}_{10}\,\mathfrak{n}(\bar{z_i}, m)}{\partial m}\bigg|_{m_{\rm lim}},
\end{equation}
where $\mathfrak{n}(\bar{z_i}, m)$ is the true distribution of galaxies, evaluated at the central redshift of bin $i$, $\bar{z_i}$. It is important to note that, in practice, this slope is determined from observations, and accordingly depends on the wavelength band within which the galaxy is observed in addition to its redshift. 

Operationally, magnification bias causes the true shear, $\gamma^\alpha_i$, to be replaced, within the estimator used to determine the power spectrum from data, by an `observed' shear:
\begin{equation}
    \label{eq:MBshear}
    \gamma^\alpha_{{\rm obs}; i} \xrightarrow{} \gamma^\alpha_i+\gamma^\alpha_i\delta^g_{{\rm obs}; i} = \gamma^\alpha_i+\gamma^\alpha_i\delta^g_i + (5s_i-2)\gamma^\alpha_i\kappa_i.
\end{equation}
Now, we can evaluate the impact of magnification bias on the two-point statistic by substituting $\widetilde{\gamma}^\alpha_{{\rm obs}; i}$ for $\widetilde{\gamma}^\alpha_i$ in Eq. (\ref{eq:Emode}), and recomputing. As source-lens clustering terms of the form $\gamma^\alpha_i\delta^g_i$ are negligible \citep{RSMBcombpap}, we recover the standard result of Eq. (\ref{eq:powerspecdef}), with an additional correction term:
\begin{align}
    \label{eq:MBEcorr}
    \delta\langle\widetilde{E}_i(\boldsymbol{\ell})\widetilde{E}_j(\boldsymbol{\ell'})\rangle &= \sum_\alpha \sum_\beta T^\alpha(\boldsymbol{\ell})T^\beta(\boldsymbol{\ell'})(5s_i-2)\braket{\widetilde{(\gamma^\alpha\kappa)}_i(\boldsymbol{\ell})\,\widetilde{\gamma}_j^\beta(\boldsymbol{\ell'})}\nonumber\\
    &+ T^\alpha(\boldsymbol{\ell'})T^\beta(\boldsymbol{\ell})(5s_j-2)\braket{\widetilde{(\gamma^\alpha\kappa)}_j(\boldsymbol{\ell'})\,\widetilde{\gamma}_i^\beta(\boldsymbol{\ell})}.
\end{align}
Analogously to the reduced shear case, we then obtain corrections to the auto and cross-correlation angular spectra of the form:
\begin{align}
    \label{eq:dClMB}
    \delta C^{\rm MB}_{\ell;ij} &= \ell(\ell+1)\frac{(\ell+2)!}{(\ell-2)!}\frac{1}{(\ell+1/2)^6}\int_0^\infty\frac{{\rm d}^2\boldsymbol{\ell'}}{(2\pi)^2}\cos(2\phi_{\ell'}-2\phi_\ell)\nonumber\\
    &\times [(5s_i-2)B_{iij}^{\kappa\kappa\kappa}(\boldsymbol{\ell}, \boldsymbol{\ell'}, -\boldsymbol{\ell}-\boldsymbol{\ell'})\nonumber\\
    &+ (5s_j-2)B_{ijj}^{\kappa\kappa\kappa}(\boldsymbol{\ell}, \boldsymbol{\ell'}, -\boldsymbol{\ell}-\boldsymbol{\ell'})].
\end{align}
We note that the mathematical form of Eq. (\ref{eq:dClMB}) is simply Eq. (\ref{eq:dCl}) with factors of $(5s_i - 2)$ and $(5s_j - 2)$ applied to the corresponding bispectra. These additional prefactors are due to the magnification bias contribution from each bin depending on the slope of the luminosity function in that bin. Accordingly, we are able to compute both of these effects for the computational cost of one.

\subsection{\label{sec:23}Intrinsic alignments}

When galaxies form near each other, they do so in a similar tidal field. Such tidal process occurring during galaxy formation, together with other processes such as spin correlations, can induce a preferred, intrinsically correlated, alignment of galaxy shapes \citep{IA1,IA2,IA3}. To first-order, this can be thought of as an additional contribution to the observed ellipticity of a galaxy, $\epsilon$:
\begin{equation}
    \label{eq:epsi1}
    \epsilon = \gamma + \gamma^{\rm I} + \epsilon^s,
\end{equation}
where $\gamma=\gamma^1+{\rm i}\gamma^2$ is the gravitational lensing shear, $\gamma^{\rm I}$ is the contribution to the observed shape resulting from IAs, and $\epsilon^s$ is the source ellipticity that the galaxy would have in the absence of the process causing the IA.

Using Eq. (\ref{eq:epsi1}), we find that the theoretical two-point statistic (e.g. the two-point correlation function, or the power spectrum) consists of three types of terms: $\braket{\gamma\gamma},\braket{\gamma^{\rm I}\gamma}$, and $\braket{\gamma^{\rm I}\gamma^{\rm I}}$. The first of these terms leads to the standard lensing power spectra of Eq. (\ref{eq:Cl}), while the other two terms lead to additional contributions to the observed power spectra, $C_{\ell;ij}^{\epsilon\epsilon}$, so that:
\begin{equation}
    \label{eq:ObsCl}
    C_{\ell;ij}^{\epsilon\epsilon} = C_{\ell;ij}^{\gamma\gamma} + C_{\ell;ij}^{{\rm I}\gamma} + C_{\ell;ij}^{\rm II} + N_{\ell;ij}^\epsilon,
\end{equation}
where $C_{\ell;ij}^{{\rm I}\gamma}$ represents the correlation between the background shear and the foreground IA, $C_{\ell;ij}^{\rm II}$ are the auto-correlation spectra of the IAs, and $N_{\ell;ij}^\epsilon$ is a shot noise term. The additional spectra can be described in a similar manner to the shear power spectra, by way of the non-linear alignment (NLA) model \citep{IA_NLA}:
\begin{align}
    \label{eq:cllig}
    C_{\ell;ij}^{{\rm I}\gamma} &= \frac{(\ell+2)!}{(\ell-2)!}\frac{1}{(\ell+1/2)^4}\int_0^{\chi_{\rm lim}}\frac{{\rm d}\chi}{d^{\,2}_{\rm A}(\chi)}[W_i(\chi)n_j(\chi) \nonumber\\
    &+n_i(\chi)W_j(\chi)] P_{\delta {\rm I}}(k, \chi),\\
    \label{eq:clli}
    C_{\ell;ij}^{\rm II} &= \frac{(\ell+2)!}{(\ell-2)!}\frac{1}{(\ell+1/2)^4}\int_0^{\chi_{\rm lim}}\frac{{\rm d}\chi}{d^{\,2}_{\rm A}(\chi)}n_i(\chi)\,n_j(\chi)\,P_{\rm II}(k, \chi),
\end{align}
where the intrinsic alignment power spectra, $P_{\delta {\rm I}}(k, \chi)$ and $P_{\rm II}(k, \chi)$, are expressed as functions of the matter power spectra:
\begin{align}
    \label{eq:pdi}
    P_{\delta {\rm I}}(k, \chi) &= \frac{-\mathcal{A}_{\rm IA}\mathcal{C}_{\rm IA}\Omega_{\rm m}}{D(\chi)}P_{\delta\delta}(k,\chi),\\
    \label{eq:pii}
    P_{\rm II}(k, \chi) &= \bigg(\frac{-\mathcal{A}_{\rm IA}\mathcal{C}_{\rm IA}\Omega_{\rm m}}{D(\chi)}\bigg)^2P_{\delta\delta}(k,\chi),
\end{align}
in which $\mathcal{A}_{\rm IA}$ and $\mathcal{C}_{\rm IA}$ are free model parameters to be determined by fitting to data or simulations, and $D(\chi)$ is the growth factor of density perturbations in the Universe, as a function of comoving distance.

\subsection{\label{sec:25}IA-enhanced lensing bias}

The reduced shear approximation is also used when considering the impact of IAs, and magnification bias plays a role here too. We account for these by substituting the appropriate second-order expansions of the shear, Eq. (\ref{eq:gexpan}) and Eq. (\ref{eq:MBshear}), in place of $\gamma$ within Eq. (\ref{eq:epsi1}). Neglecting source-lens clustering, the ellipticity now becomes:
\begin{equation}
    \label{eq:epsi2}
    \epsilon \simeq \gamma + (1+5s-2)\gamma\kappa + \gamma^{\rm I} + \epsilon^s.
\end{equation}
Constructing a theoretical expression for the two-point statistic from this revised expression for the ellipticity now gives us six types of terms: $\braket{\gamma\gamma}, \braket{\gamma^{\rm I}\gamma}, \braket{\gamma^{\rm I}\gamma^{\rm I}}, \braket{(\gamma\kappa)\gamma}, \braket{(\gamma\kappa)(\gamma\kappa)},$ and $\braket{(\gamma\kappa)\gamma^{\rm I}}$. The first three terms remain unchanged from the first-order case. The fourth term encompasses the basic reduced shear and magnification bias corrections, and results in the shear power spectrum corrections defined by Eq. (\ref{eq:dCl}) and Eq. (\ref{eq:dClMB}). The fifth of these terms can be neglected, as it is a fourth-order term. The final term creates an additional correction, $\delta C_{\ell;ij}^{\rm I}$, to the observed spectra that takes a form analogous to the basic reduced shear and magnification bias corrections:
\begin{align}
    \label{eq:dClIA}
    \delta C_{\ell;ij}^{\rm I} &= \ell(\ell+1)\frac{(\ell+2)!}{(\ell-2)!}\frac{1}{(\ell+1/2)^6}\int_0^\infty\frac{{\rm d}^2\boldsymbol{\ell'}}{(2\pi)^2}\cos(2\phi_{\ell'}) \nonumber\\
    &\times [(1+5s_i-2)B_{iij}^{\kappa\kappa {\rm I}}(\boldsymbol{\ell}, \boldsymbol{\ell'}, -\boldsymbol{\ell}-\boldsymbol{\ell'})\nonumber\\
    &+ (1+5s_j-2)B_{jji}^{\kappa\kappa {\rm I}}(\boldsymbol{\ell}, \boldsymbol{\ell'}, -\boldsymbol{\ell}-\boldsymbol{\ell'})],
\end{align}
where the convergence-IA bispectra, $B_{iij}^{\kappa\kappa {\rm I}}$ and $B_{jji}^{\kappa\kappa {\rm I}}$, are given by:
\begin{align}
    \label{eq:scIBi}
    B_{iij}^{\kappa\kappa {\rm I}}(\boldsymbol{\ell_1}, \boldsymbol{\ell_2}, \boldsymbol{\ell_3}) &= \int_0^{\chi_{\rm lim}}\frac{{\rm d}\chi}{d^{\,4}_{\rm A}(\chi)}W^2_i(\chi)n_j(\chi)B_{\delta\delta {\rm I}}(\boldsymbol{k_1},\boldsymbol{k_2},\boldsymbol{k_3},\chi),\hspace{-0.1mm}\\
    B_{jji}^{\kappa\kappa {\rm I}}(\boldsymbol{\ell_1}, \boldsymbol{\ell_2}, \boldsymbol{\ell_3}) &= \int_0^{\chi_{\rm lim}}\frac{{\rm d}\chi}{d^{\,4}_{\rm A}(\chi)}W^2_j(\chi)n_i(\chi)B_{\delta\delta {\rm I}}(\boldsymbol{k_1},\boldsymbol{k_2},\boldsymbol{k_3},\chi).
\end{align}
The density perturbation-IA bispectrum, $B_{\delta\delta {\rm I}}(\boldsymbol{k_1},\boldsymbol{k_2},\boldsymbol{k_3},\chi)$, can be calculated in a similar way to the matter density perturbation bispectrum, using perturbation theory and the \cite{Scoccimarro01} fitting formula. Accordingly:
\begin{align}
    \label{eq:bispecEN}
    B_{\delta\delta {\rm I}}(\boldsymbol{k_1},\boldsymbol{k_2},\boldsymbol{k_3},\chi) &= 2F_2^{\rm eff}(\boldsymbol{k_1},\boldsymbol{k_2}) P_{{\rm I}\delta}(k_1, \chi)P_{\delta\delta}(k_2, \chi) \nonumber\\
    &+ 2F_2^{\rm eff}(\boldsymbol{k_2},\boldsymbol{k_3})P_{\delta\delta}(k_2, \chi)P_{\delta {\rm I}}(k_3, \chi)  \nonumber\\
    &+ 2F_2^{\rm eff}(\boldsymbol{k_1},\boldsymbol{k_3})P_{\delta {\rm I}}(k_1, \chi)P_{\delta\delta}(k_3, \chi),
\end{align}
with $P_{{\rm I}\delta}(k_1, \chi)=P_{\delta {\rm I}}(k_1, \chi)$. This equation is an ansatz for how IAs behave in the non-linear regime, analogous to the NLA model. The described approach, and in particular the fitting functions, remain valid because, in the NLA model, we can treat IAs as a field proportional, by some redshift-dependence weighting, to the matter density contrast. Since the fitting functions, $F_2^{\rm eff}$, do not depend on the comoving distance, they remain unchanged. For the full derivation of this bispectrum term, and a generalisation for similar terms, see Appendix \ref{AGB}. 

\subsection{\label{sec:24}Shot noise}

The shot noise term in Eq. (\ref{eq:ObsCl}) arises from the uncorrelated part of the unlensed source ellipticities; represented by $\epsilon^s$ in Eq. (\ref{eq:epsi1}). This is zero for cross-correlation spectra, because the ellipticities of galaxies at different comoving distances should be uncorrelated. However, it is non-zero for auto-correlation spectra. It is written as:
\begin{equation}
    \label{eq:shotnoise}
    N_{\ell;ij}^\epsilon = \frac{\sigma_\epsilon^2}{\bar{n}_{\rm g}/N_{\rm bin}}\delta_{ij}^{\rm K},
\end{equation}
where $\sigma_\epsilon^2$ is the variance of the observed ellipticities in the galaxy sample, $\bar{n}_{\rm g}$ is the galaxy surface density of the survey, $N_{\rm bin}$ is the number of tomographic bins used, and $\delta_{ij}^{\rm K}$ is the Kronecker delta. Equation (\ref{eq:shotnoise}) assumes the bins are equi-populated.

\subsection{\label{sec:26}Fisher and bias formalism}

To estimate the uncertainty on cosmological parameters that will be obtained from \emph{Euclid}, we used the Fisher matrix approach \citep{Tegmark97, ISTforecast}. In this formalism, the Fisher matrix is defined as the expectation of the Hessian of the likelihood:
\begin{equation}
    \label{eq:fishbasic}
    F_{\tau\zeta} = \bigg\langle\frac{-\partial^2 \ln L}{\partial\theta_\tau\partial\theta_\zeta}\bigg\rangle,
\end{equation}
where $L$ is the likelihood of the parameters given the data, and $\tau$ and $\zeta$ refer to parameters of interest, $\theta_\tau$ and $\theta_\zeta$. Assuming a Gaussian likelihood, the Fisher matrix can be rewritten in terms of only the covariance of the data, $\boldsymbol{C}$, and the mean of the data vector, $\boldsymbol{\mu}$:
\begin{equation}
    \label{eq:fishgauss}
    F_{\tau\zeta} = \frac{1}{2}\:{\rm tr}\,\bigg[\frac{\partial\boldsymbol{C}}{\partial\theta_\tau}\boldsymbol{C}^{-1}\frac{\partial\boldsymbol{C}}{\partial\theta_\zeta}\boldsymbol{C}^{-1}\bigg] + \sum_{pq}\frac{\partial\mu_p}{\partial\theta_\tau}(\boldsymbol{C}^{-1})_{pq}\frac{\partial\mu_q}{\partial\theta_\zeta},
\end{equation}
where the summations over $p$ and $q$ are summations over the variables in the data vector. In the case of cosmic shear, we can take our signal to be the mean of the power spectrum, so the first term in Eq. (\ref{eq:fishgauss}) vanishes.

In reality, the weak lensing likelihood is non-Gaussian (see e.g. \cite{nonGaussLik}). However, recent investigations indicate that the assumption of a Gaussian likelihood is unlikely to lead to significant biases in the cosmological parameters inferred from a Stage IV weak lensing experiment \citep{LSSTnonGauss, 2019arXiv190405364T}. Additionally, while this non-Gaussianity affects the shapes of the constraints on cosmological parameters, it does not affect the calculation of the reduced shear and magnification bias corrections, and accordingly does not significantly affect the corresponding relative biases. For these reasons, coupled with its simplicity, we proceed under the Gaussian likelihood assumption for this work.

Similarly, the covariance of the data itself is non-Gaussian. But, in contrast with the likelihood, we cannot assume a Gaussian covariance for cosmic shear (see e.g. \cite{WLSSCpap, HuSSC}. The dominant contribution to the non-Gaussian part of the covariance is the super-sample covariance (SSC) \citep{SSCorig}. This additional component arises from the fact that, in any galaxy survey, a limited fraction of the Universe is observed. Density fluctuations with wavelengths larger than the size of the survey can then cause the background density measured by the survey to no longer be representative of the true average density of the Universe. Additional non-Gaussian contributions, such as connected trispectrum terms, can be safely neglected for \emph{Euclid} \citep{WLSSCpap}.

For weak lensing, the covariance can then be expressed as the sum of the Gaussian, ${\rm Cov}_{\rm G}$, and SSC, ${\rm Cov}_{\rm SSC}$, parts:
\begin{align}
    \label{eq:covsum}
    {\rm Cov}\left[C^{\epsilon\epsilon}_{\ell;ij},C^{\epsilon\epsilon}_{\ell^\prime;mn}\right] &= {\rm Cov}_{\rm G}\left[C^{\epsilon\epsilon}_{\ell;ij},C^{\epsilon\epsilon}_{\ell^\prime;mn}\right]\nonumber\\
    &+ {\rm Cov}_{\rm SSC}\left[C^{\epsilon\epsilon}_{\ell;ij},C^{\epsilon\epsilon}_{\ell^\prime;mn}\right],
\end{align}
where $(i,j)$ and $(m,n)$ are redshift bin pairs. The Gaussian covariance is given by:
\begin{align}
    \label{eq:gauscov}
    {\rm Cov}_{\rm G}\left[C^{\epsilon\epsilon}_{\ell;ij},C^{\epsilon\epsilon}_{\ell^\prime;mn}\right] &=
\frac{C^{\epsilon\epsilon}_{\ell;i m}\,C^{\epsilon\epsilon}_{\ell^\prime;jn}+C^{\epsilon\epsilon}_{\ell;in}\,C^{\
\epsilon\epsilon}_{\ell^\prime;jm}}{(2 \ell + 1) f_{\rm sky} \Delta \ell}\,\delta^{\rm K}_{\ell\ell^\prime},
\end{align}
where $\delta^{\rm K}$ is the Kronecker delta, $\Delta\ell$ is the bandwidth of $\ell$-modes sampled, and $f_{\rm sky}$ is the fraction of the sky surveyed.The contribution from SSC can be approximated as \citep{fastssc}:
\begin{equation}
    \label{eq:covssc}
    {\rm Cov}_{\rm SSC}\left[C^{\epsilon\epsilon}_{\ell;ij},C^{\epsilon\epsilon}_{\ell^\prime;mn}\right] \approx R_\ell\, C^{\epsilon\epsilon}_{\ell;ij}\, R_\ell\, C^{\epsilon\epsilon}_{\ell^\prime;mn}\,S_{ijmn},
\end{equation}
where $S_{ijmn}$ is the dimensionless volume-averaged covariance of the background matter density contrast, and $R_\ell$ is the effective relative response of the observed power spectrum.

The specific Fisher matrix used in this investigation can be expressed as:
\begin{equation}
    \label{eq:fishGauss}
    F^{\rm G}_{\tau\zeta} = \sum_{\ell'=\ell_{\rm min}}^{\ell_{\rm max}}\sum_{\ell=\ell_{\rm min}}^{\ell_{\rm max}}\sum_{ij,mn}
             \frac{\partial C^{\epsilon\epsilon}_{\ell;ij}}{\partial \theta_{\tau}} {\rm Cov}^{-1}\left[C^{\
\epsilon\epsilon}_{\ell;ij},C^{\epsilon\epsilon}_{\ell';mn}\right]
\frac{\partial C^{\epsilon\epsilon}_{\ell';mn}}{\partial \theta_{\zeta}},
\end{equation}
where $(\ell_{\rm min}, \ell_{\rm max})$ are the minimum and maximum angular wavenumbers used, and the sum is over the $\ell$-blocks. 

From this, we can calculate the expected uncertainties on our parameters, $\sigma_\tau$, using the relation:
\begin{equation}
    \label{eq:sigma}
    \sigma_\tau = \sqrt{({F^{-1}})_{\tau\tau}}.
\end{equation}
The Fisher matrix can also be used to determine the projected confidence region ellipses of pairs of cosmological parameters \citep{ISTforecast}.

In the presence of a systematic effect in the signal, the Fisher matrix formalism can be adapted to measure how biased the inferred cosmological parameter values are if this effect is not taken into consideration \citep{biaspap}. This bias is calculated as follows:
\begin{align}
    \label{eq:bias}
    b(\theta_\tau) &= \sum_\zeta{(F^{-1})}_{\tau\zeta}\: B_\zeta,
\end{align}
with:
\begin{align}
    \label{eq:sscbias}
    B_{\zeta} &=
\sum_{\ell'=\ell_{\rm min}}^{\ell_{\rm max}}\sum_{\ell=\ell_{\rm min}}^{\ell_{\rm max}}\sum_{ij,mn}\delta C_{\ell;ij}\,
{\rm Cov}^{-1}\left[C^{\
\epsilon\epsilon}_{\ell;ij},C^{\epsilon\epsilon}_{\ell';mn}\right]\;\frac{\partial C_{\ell';mn}}{\partial \zeta},
\end{align}
where $\delta C_{\ell;ij}$ is the value of the systematic effect for bins $(i,j)$, in our case reduced shear and magnification bias.

\section{\label{sec:3}Methodology}

In order to quantify the impact of the three corrections on \emph{Euclid}, we adopted the forecasting specifications of \cite{ISTforecast}. Accordingly, we took there to be ten equi-populated tomographic bins, with bin edges: \{0.001, 0.418, 0.560, 0.678, 0.789, 0.900, 1.019, 1.155, 1.324, 1.576, 2.50\}.
\begin{table}[b]
\centering
\caption{Fiducial values of $w$CDM cosmological parameters for which the bias from reduced shear and magnification bias corrections is calculated. These values were selected in accordance with \emph{Euclid} Collaboration forecasting choices \citep{ISTforecast}; to facilitate consistent comparisons. We note that the value of the neutrino mass was kept fixed in the Fisher matrix calculations.}
\label{tab:cosmology}
\begin{tabular}{c c}
\hline\hline
Cosmological Parameter & Fiducial Value\\
\hline
$\Omega_{\rm m}$ & 0.32 \\
$\Omega_{\rm b}$ & 0.05 \\
$h$ & 0.67 \\
$n_{\rm s}$ & 0.96 \\
$\sigma_8$ & 0.816 \\
$\sum m_\nu$ (eV) & 0.06 \\
$\Omega_{\rm DE}$ & 0.68 \\
$w_0$ & $-1$ \\
$w_a$ & 0  \\
\hline\hline
\end{tabular}
\end{table}
We primarily investigated the impact on the `optimistic' case for such a survey, in which $\ell$-modes of up to 5000 are probed, because this is necessary for \emph{Euclid} to reach its required figure of merit using cosmic shear \citep{ISTforecast}. For the `pessimistic' case, see Appendix \ref{A:Pess}. We considered the intrinsic variance of observed ellipticities to have two components, each with a value of 0.21, so that the intrinsic ellipticity root-mean-square value $\sigma_\epsilon = \sqrt{2}\times0.21 \approx 0.3$. For \emph{Euclid}, we took the surface density of galaxies to be $\bar{n}_{\rm g}=30$ arcmin$^{-2}$, and the fraction of sky covered to be $f_{\rm sky}=0.36$.

Furthermore, we considered the $w$CDM model case in our calculations. This extension of the $\Lambda$CDM model accounts for a time-varying dark energy equation of state. In this model, we have the following parameters: the present-day matter density parameter $\Omega_{\rm m}$, the present-day baryonic matter density parameter $\Omega_{\rm b}$, the Hubble parameter $h=H_0/100$km\:s$^{-1}$Mpc$^{-1}$, the spectral index $n_{\rm s}$, the RMS value of density fluctuations on 8 $h^{-1}$Mpc scales $\sigma_8$, the present-day dark energy density parameter $\Omega_{\rm DE}$, the present-day value of the dark energy equation of state $w_0$, and the high redshift value of the dark energy equation of state $w_a$. Additionally, we assumed neutrinos to have masses. We denote the sum of neutrino masses by $\sum m_\nu\ne 0$. This quantity was kept fixed, and we did not generate confidence contours for it, in concordance with \cite{ISTforecast}. The fiducial values chosen for these parameters are given in Table \ref{tab:cosmology}. These values were chosen to allow for a direct and consistent comparison of the two corrections with the forecasted precision of \emph{Euclid}. The values provided in the forecasting specifications for the free parameters of the NLA model were also used in our work, in Eq. (\ref{eq:pdi}) and Eq. (\ref{eq:pii}). These are: $\mathcal{A}_{\rm IA}=1.72$ and $\mathcal{C}_{\rm IA}=0.0134$.
\begin{table}[t]
    \centering
    \caption{Choice of parameter values used to define the probability distribution function of the photometric redshift distribution of sources, in Eq. (\ref{eq:pphot}). We did not consider how variation in the quality of photometric redshifts impacts the Fisher matrix predictions.}
    \begin{tabular}{c c}
    \hline\hline
    Model Parameter& Fiducial Value\\
    \hline
    $c_{\rm b}$ & 1.0\\
    $z_{\rm b}$ & 0.0\\
    $\sigma_{\rm b}$ & 0.05\\
    $c_{\rm o}$ & 1.0\\
    $z_{\rm o}$ & 0.1\\
    $\sigma_{\rm o}$ & 0.05\\
    $f_{\rm out}$ & 0.1\\
    \hline\hline
    \end{tabular}
    \label{tab:phphotparams}
\end{table}

As in \cite{ISTforecast}, we chose to define the distributions of galaxies in our tomographic bins, for photometric redshift estimates, as:
\begin{equation}
    \label{eq:ncfht}
    {\mathcal N}_i(z) = \frac{\int_{z_i^-}^{z_i^+}{\rm d}z_{\rm p}\,\mathfrak{n}(z)p_{\rm ph}(z_{\rm p}|z)}{\int_{z_{\rm min}}^{z_{\rm max}}{\rm d}z\int_{z_i^-}^{z_i^+}{\rm d}z_{\rm p}\,\mathfrak{n}(z)p_{\rm ph}(z_{\rm p}|z)},
\end{equation}
where $z_{\rm p}$ is measured photometric redshift, $z_i^-$ and $z_i^+$ are edges of the $i$-th redshift bin, and $z_{\rm min}$ and $z_{\rm max}$ define the range of redshifts covered by the survey. Then, $n_i(\chi) = {\mathcal N}_i(z){\rm d}z/{\rm d}\chi$. In Eq. (\ref{eq:ncfht}), $\mathfrak{n}(z)$ is the true distribution of galaxies with redshift, $z$; defined as in the \textit{Euclid} Red Book \citep{EuclidRB}:
\begin{equation}
    \label{eq:ntrue}
    \mathfrak{n}(z) \propto \bigg(\frac{z}{z_0}\bigg)^2\,{\rm exp}\bigg[-\bigg(\frac{z}{z_0}\bigg)^{3/2}\bigg],
\end{equation}
where $z_0=z_{\rm m}/\sqrt{2}$, with $z_{\rm m}=0.9$ as the median redshift of the survey. Meanwhile, the function $p_{\rm ph}(z_{\rm p}|z)$ describes the probability that a galaxy at redshift $z$ is measured to have a redshift $z_{\rm p}$, and takes the parameterisation:
\begin{align}
\label{eq:pphot}
        p_{\rm ph}(z_{\rm p}|z) &= \frac{1-f_{\rm out}}{\sqrt{2\pi}\sigma_{\rm b}(1+z)}\,{\rm exp}\Bigg\{-\frac{1}{2}\bigg[\frac{z-c_{\rm b}z_{\rm p}-z_{\rm b}}{\sigma_{\rm b}(1+z)}\bigg]^2\Bigg\} \nonumber\\
        &+ \frac{f_{\rm out}}{\sqrt{2\pi}\sigma_{\rm o}(1+z)}\:{\rm exp}\Bigg\{-\frac{1}{2}\bigg[\frac{z-c_{\rm o}z_{\rm p}-z_{\rm o}}{\sigma_{\rm o}(1+z)}\bigg]^2\Bigg\}.
\end{align}
In this parameterisation, the first term describes the multiplicative and additive bias in redshift determination for the fraction of sources with a well measured redshift, whereas the second term accounts for the effect of a fraction of catastrophic outliers, $f_{\rm out}$. The values of these parameters, chosen to match the selection of \cite{ISTforecast}, are stated in Table \ref{tab:phphotparams}. By using this formalism, the impact of the photometric redshift uncertainties was also included in the derivatives, with respect to the cosmological parameters, of the shear power spectra.

The matter density power spectrum and growth factor used in our analyses were computed using the publicly available \texttt{CLASS}\footnote{\url{https://class-code.net/}} cosmology package \citep{Classpap}. Within the framework of \texttt{CLASS}, we included non-linear corrections to the matter density power spectrum, using the \texttt{Halofit} model \citep{Takahashi12}. Using these modelling specifics, we first calculated the basic reduced shear correction of Eq. (\ref{eq:dCl}), and the resulting biases in the $w$CDM parameters. In doing so, we computed the derivatives of our tomographic matrices, at each sampled $\ell$-mode, using a simple finite-difference method. To calculate the dimensionless volume-averaged covariance of the background matter density contrast of Eq. \ref{eq:covssc}, we used the publicly available \texttt{PySSC}\footnote{\url{https://github.com/fabienlacasa/PySSC}} code \citep{fastssc} to compute the full-sky value, and divided by the \emph{Euclid} value of $f_{\rm sky}$. Additionally, we set $R_\ell \approx 4$ for weak lensing\footnote{Private communications with F. Lacasa.}.

Our Fisher matrices included the parameters $\Omega_{\rm m}, \Omega_{\rm b}, h, n_{\rm s}, \sigma_8, \Omega_{\rm DE}, w_0, w_a,$ and $\mathcal{A}_{\rm IA}$. We did not include any additional nuisance parameters.  However, we do not expect this to affect the significance of the corrections, as \cite{ISTforecast} find that the inclusion of various nuisance parameters typically alters the predicted relative uncertainties on cosmological parameters by less than $10\%$. No prior was added to our Fisher matrix analysis.

The correction for magnification bias, and the resulting biases in the cosmological parameters, were calculated in the same way. 
\begin{table}[t]
    \centering
    \caption{Slope of the luminosity function for each redshift bin, calculated at the central redshifts of each bin. These are evaluated at the limiting magnitude 24.5 (AB in the \emph{Euclid} VIS band \citep{VISpap}). The slopes are determined using finite difference methods with the fitting formula of \cite{JoachimiBridleMB}, which is based on fitting to COMBO-17 and SDSS $r$-band results \citep{BlakeBridlenz}.} 
    \begin{tabular}{c c c}
    \hline\hline
    Bin $i$ & Central Redshift & Slope $s_i$\\
    \hline
    1 & 0.2095 & 0.196\\
    2 & 0.489 & 0.274\\
    3 & 0.619 & 0.320\\
    4 & 0.7335 & 0.365\\
    5 & 0.8445 & 0.412\\
    6 & 0.9595 & 0.464\\
    7 & 1.087 & 0.525\\
    8 & 1.2395 & 0.603\\
    9 & 1.45 & 0.720\\
    10 & 2.038 & 1.089\\
    \hline\hline
    \end{tabular}
    \label{tab:slopesbybin}
\end{table}
The slope of the luminosity function, as defined in Eq. (\ref{eq:slope}), was calculated for each redshift bin using the approach described in Appendix C of \cite{JoachimiBridleMB}. We applied a finite-difference method to the fitting formula for galaxy number density as a function of limiting magnitude stated here, in order to calculate the slope of the luminosity function at the limiting magnitude of \emph{Euclid}, 24.5 \citep{EuclidRB}; or AB in the \emph{Euclid} VIS band \citep{VISpap}. This technique produces slope values consistent with those generated from the Schechter function approach of \cite{LSSTMagBias}. The calculated slopes for each redshift bin are given in Table \ref{tab:slopesbybin}. However, we emphasise that while this method allows the investigation of the impact of magnification bias at this stage, when the correction is computed for the true \emph{Euclid} data, updated galaxy number counts determined directly from \emph{Euclid} observations should be used to ensure accuracy.

We then combined the two corrections, and calculated the resulting biases, as well as the resulting confidence contours for parameter combinations. Next, the additional IA-lensing bias interaction term from Eq. (\ref{eq:dClIA}) was included, and the biases were recomputed.

To validate the perturbative formalism based on a fitting formula for the matter bispectrum, we also computed the reduced shear correction using a forward model approach assuming the lognormal field approximation~\citep{hilbert2011cosmic, mancini20183d, xavier2016improving}. This approximation was recently used to generate a covariance matrix in the Dark Energy Survey Year 1 analysis~\citep{troxel2018dark}. Using the pipeline recently presented in~\cite{2019arXiv190405364T} (which uses the public code {\tt CAMB}~\citep{lewis538efficient}, {\tt Halofit}~\citep{Takahashi12}, {\tt Cosmosis}~\citep{zuntz2015cosmosis} and the python wrapper of {\tt HEALpix}~\footnote{\url{https://sourceforge.net/projects/healpix/}}~\citep{gorski1999healpix, 2005ApJ...622..759G} -- {\tt HEALpy}) we computed the reduced shear correction by averaging over $100$ forward realisations. We compared our semi-analytic approach to the forward modelled approach, for the auto-correlation spectrum of a single tomographic bin spanning the entire redshift range of 0 -- 2.5. To ensure a consistent comparison was made with the forward model approach, the correction computed from the perturbative formalism in this case used the best-fitting photometric redshift galaxy distribution of the CFHTLenS catalogue \citep{cfhtnpap}:
\begin{equation}
    \label{eq:nfm}
        \mathfrak{n}(z) = 1.5\,{\rm exp}\bigg[-\frac{(z-0.7)^2}{0.1024}\bigg] + 0.2\,{\rm exp}\bigg[-\frac{(z-1.2)^2}{0.2116}\bigg],
\end{equation}
as this is used in \cite{2019arXiv190405364T}. In this comparison, we did not consider magnification bias, or the IA-enhanced lensing bias case.

\section{\label{sec:4}Results and discussion}

In this section, we report the impact of the various effects studied on \emph{Euclid}. We first present the individual and combined impacts of the reduced shear and magnification bias corrections. The impact of IA-enhanced lensing bias is also discussed. Finally, we present a forward modelled approach for computing the reduced shear correction.

\subsection{\label{subsec:RSres}The reduced shear correction}
\begin{figure*}[t]
\centering
     \subfloat{%
      \includegraphics[width=0.47\linewidth]{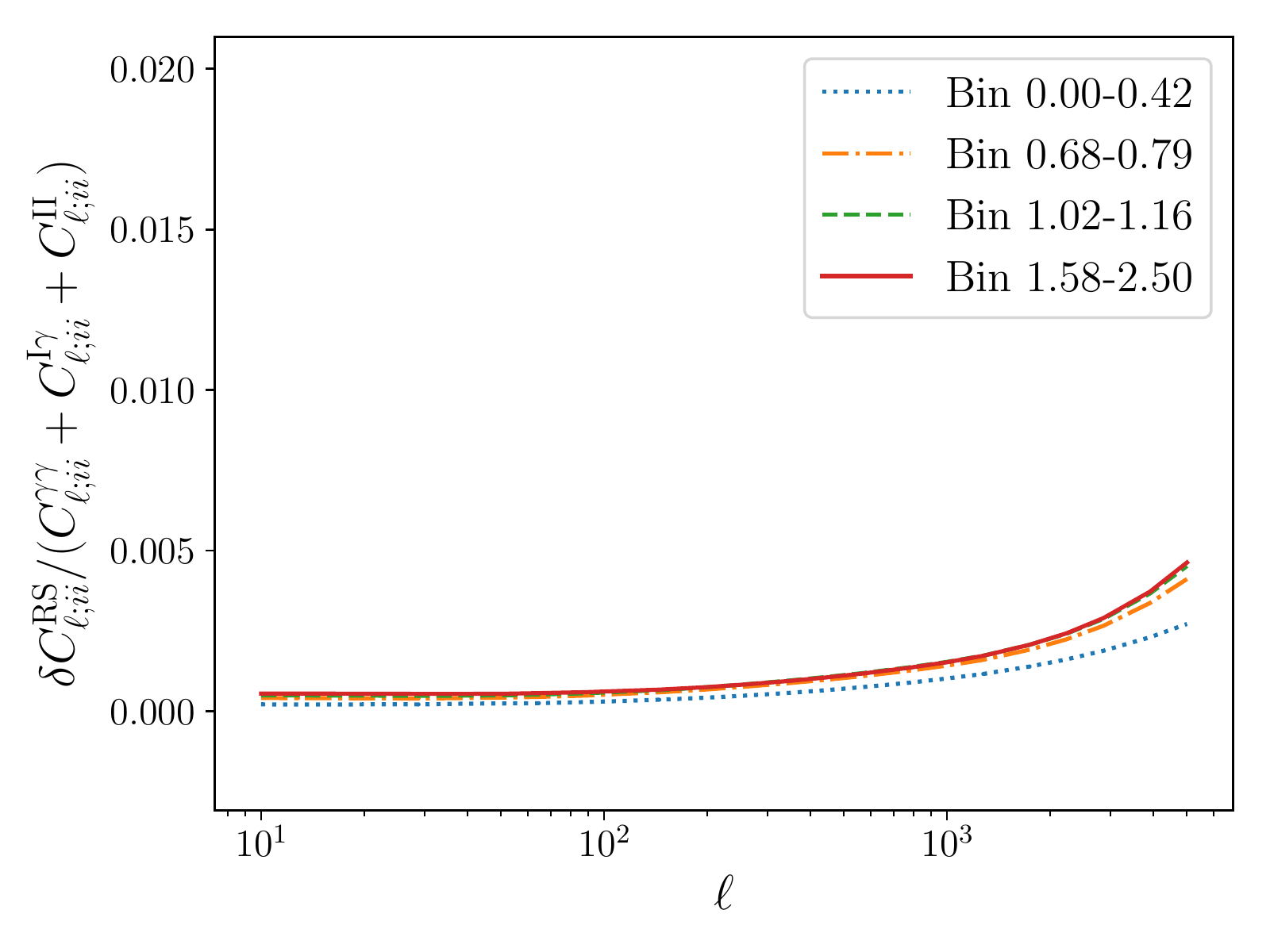}
     }
     \hfill
     \subfloat{%
      \includegraphics[width=0.47\linewidth]{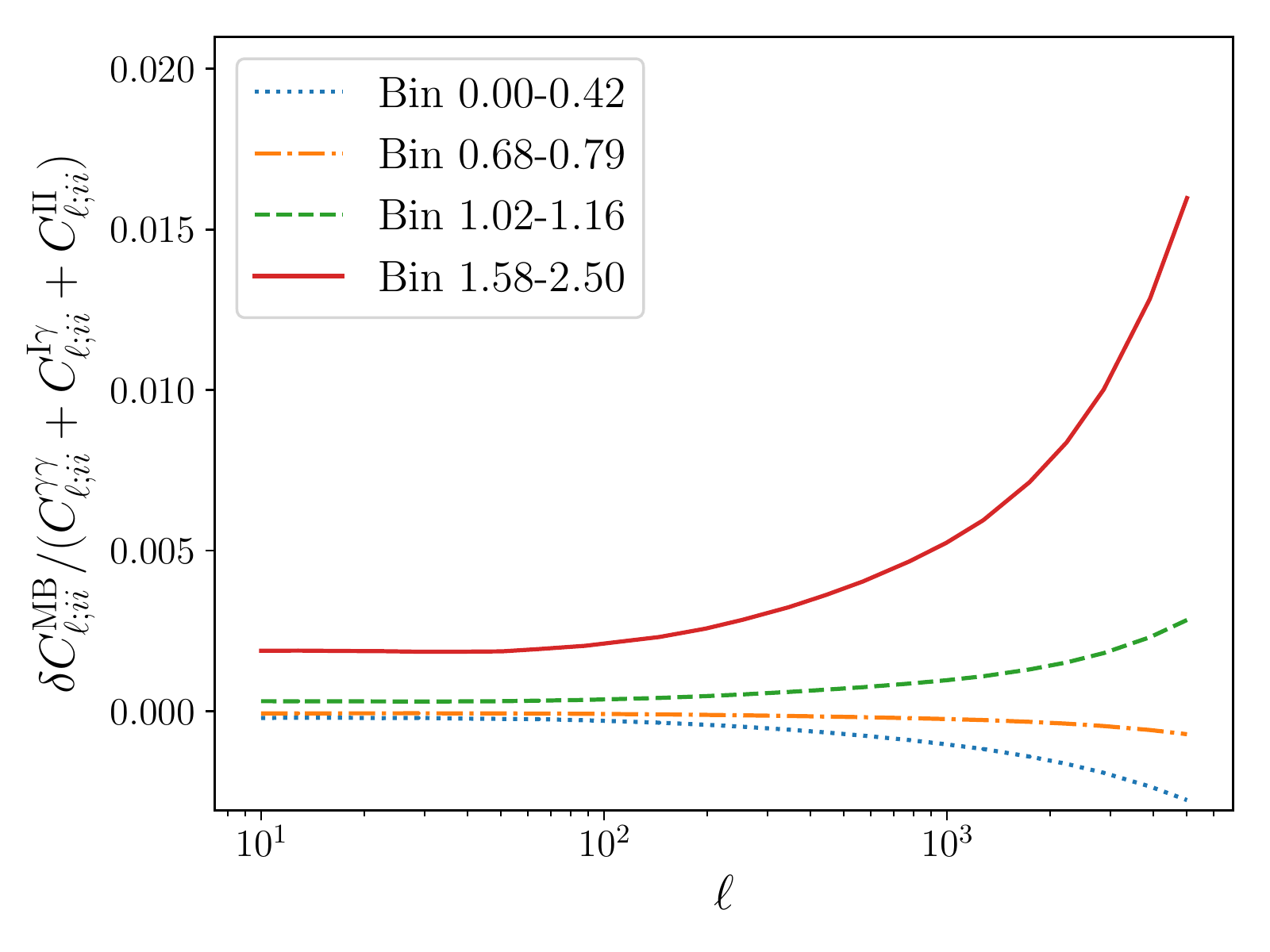}
     }
     
     \subfloat{%
      \includegraphics[width=0.47\linewidth]{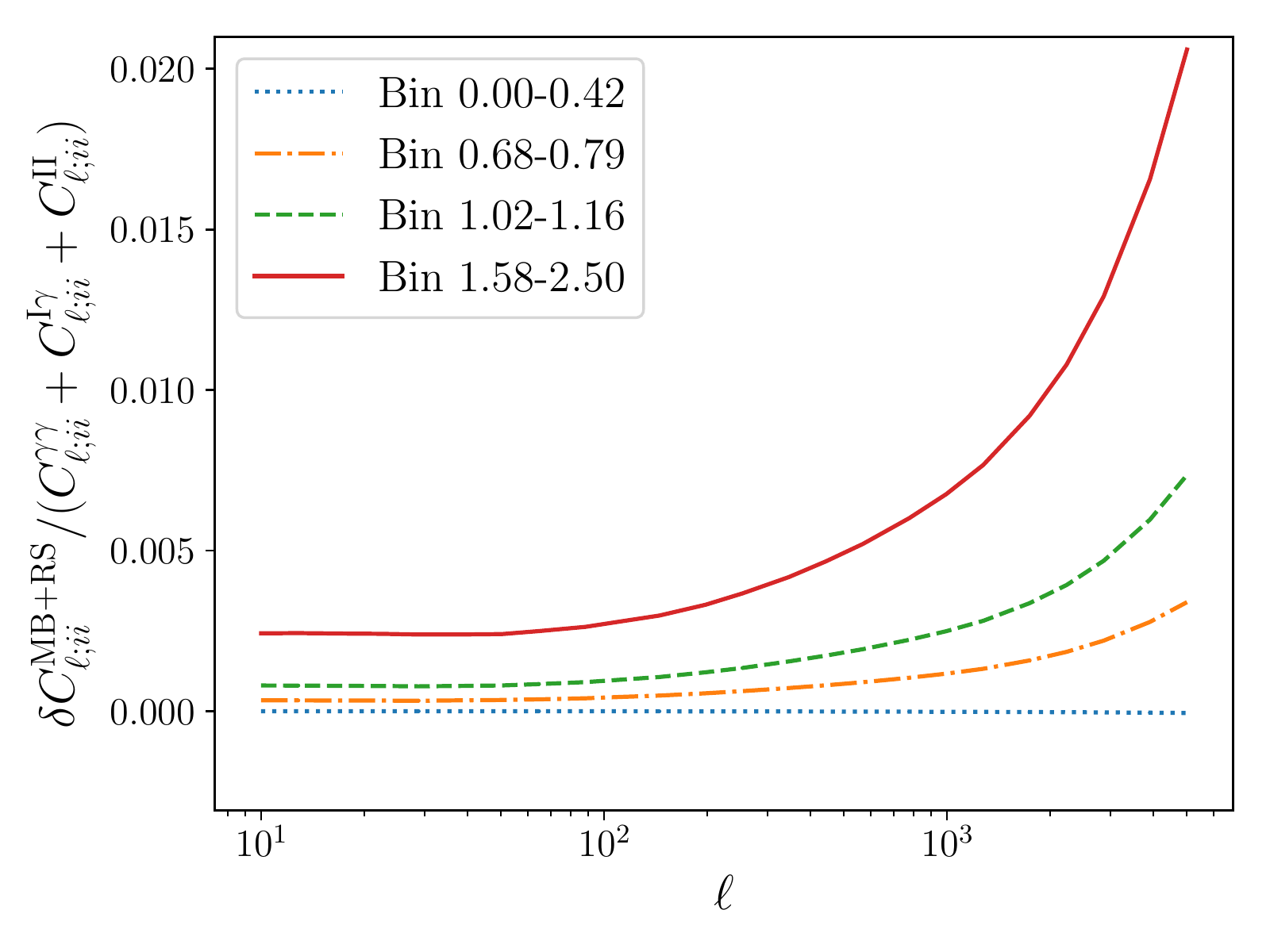}
     }
\caption{\label{fig:Comps_dcl} Reduced shear (top left), magnification bias (top right), and combined (bottom) corrections relative to the observed angular shear auto-correlation spectra (excluding shot noise), for four different redshift bins. For the basic reduced shear correction, the relative size of the correction increases alongside redshift, as the correction term has an additional factor of the lensing kernel compared to the power spectra. The correction plateaus at higher redshifts, because the lensed light encounters the most non-linearity and clustering at lower redshifts. It also increases with $\ell$, as convergence tends to be higher on smaller physical scales. For the basic magnification bias correction, the relative size of the correction also increases with redshift. At lower redshifts, the term is subtractive, as the magnification of individual galaxies dominates, leading to an overestimation of the galaxy density. Whereas, at higher redshifts, the dilution of galaxy density dominates, leading to an underestimation of the power spectra if the correction is not made. For the combined effect of the two corrections, the magnification bias correction effectively cancels out the reduced shear correction at the lowest redshifts. Meanwhile, at intermediate redshifts, the magnification bias is small, but additive; slightly enhancing the reduced shear correction. However, at the highest redshifts, the magnification bias is particularly strong, and the combined correction is significantly greater than at lower redshifts. The corrections seen here are in the case of the $w$CDM cosmology of Table \ref{tab:cosmology}.}
\end{figure*}

The relative magnitude of the basic reduced shear correction described by Eq. (\ref{eq:dCl}), to the observed shear auto-correlation spectra (excluding shot noise), at various redshifts, is shown in Fig. \ref{fig:Comps_dcl}. The correction increases with $\ell$, and becomes particularly pronounced at scales above $\ell\sim 100$. This is expected, as small-scale modes grow faster in high-density regions, where the convergence tends to be greater, so there is more power in these regions. We can also see, from Fig. \ref{fig:Comps_dcl}, that the relative magnitude of the correction increases with redshift, as the reduced shear correction has an extra factor of the lensing kernel, $W_i(\chi)$, in comparison to the angular shear spectra. The lensing kernel increases with comoving distance and, accordingly, redshift. While only a selection of auto-correlation spectra are presented in Fig. \ref{fig:Comps_dcl} for illustration purposes, the remaining auto and cross-correlation spectra exhibit the same trends. 

The uncertainties on the $w$CDM cosmological parameters that are predicted for \emph{Euclid}, are stated in Table \ref{tab:unctab}. Correspondingly, Table \ref{tab:nonIAbiastab} shows the biases that are induced in the predicted cosmological parameters from neglecting the basic reduced shear correction.

\begin{figure*}[t]
\centering
\includegraphics[width=\textwidth]{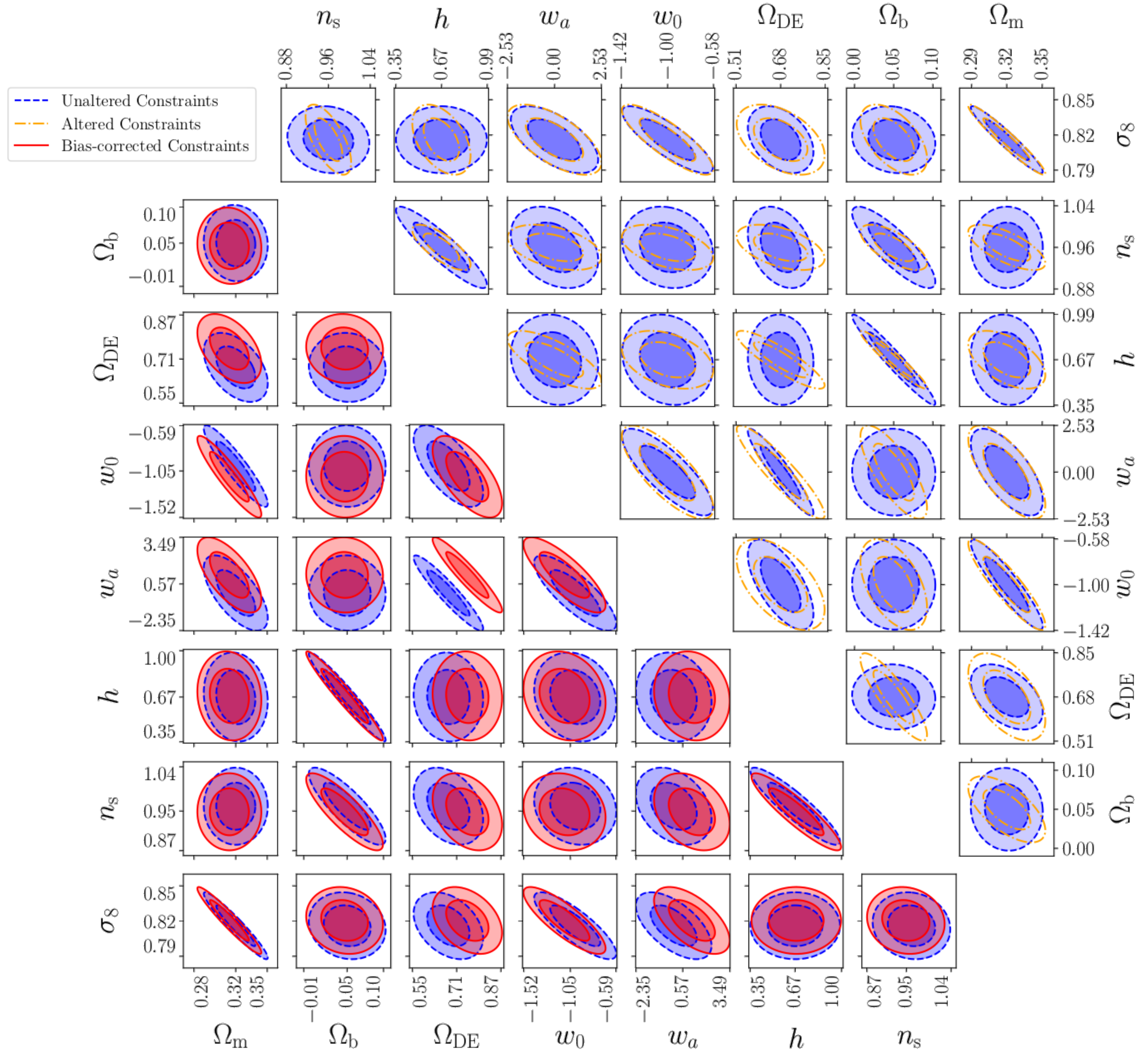}
\caption[width=2.0\linewidth]{Predicted 2-parameter projected 1-$\sigma$ and 2-$\sigma$ contours on the $w$CDM cosmological parameters from Table \ref{tab:cosmology}, for \emph{Euclid}. The optimistic case, probing $\ell$-modes up to 5000, is considered here. The biases in the predicted values of the cosmological parameters, that arise from neglecting the basic reduced shear and magnification bias corrections, are shown here (bottom left). The additional IA-lensing bias terms are not included. Of these, $\Omega_{\rm m}$, $\sigma_{8}$, $n_{\rm s}$, $\Omega_{\rm DE}$, $w_0$, and $w_a$ have significant biases of $-0.53\sigma$, $0.43\sigma$, $-0.34\sigma$, $1.36\sigma$, $-0.68\sigma$, and $1.21\sigma$, respectively. Additionally, the altered contours from including the reduced shear and magnification bias corrections, and their derivatives, in the Fisher matrix calculation are also shown (top right). The contours decrease in size for the parameters $\Omega_{\rm b}$, $h$, $n_{\rm s}$, $w_0$, and $w_a$. However, in the case of $\Omega_{\rm DE}$, the contours increase in size.}
\label{fig:bias_nonIA}
\end{figure*}
Biases are typically considered acceptable when the biased and unbiased confidence regions have an overlap of at least 90$\%$; corresponding to the magnitude of the bias being $<$ $0.25\sigma$ \citep{BiasAceptPap}. The majority of the biases are not significant, with $\Omega_{\rm m}$, $\Omega_{\rm b}$, $h$, $n_{\rm s}$, and $\sigma_8$ remaining strongly consistent pre- and post-correction. However, $\Omega_{\rm DE}$, $w_0$, and $w_a$, all exhibit significant biases of $0.31\sigma$, $-0.32\sigma$, and $0.39\sigma$, respectively. Since one of the chief goals of upcoming weak lensing surveys is the inference of dark energy parameters, these biases, which can be readily dealt with, indicate that the reduced shear correction must be included when constraining cosmological parameters from the surveys. Also shown in Table \ref{tab:unctab} is the change in the uncertainty itself, when the reduced shear correction and its derivatives are included in the Fisher matrix used for prediction. In general, the change is negligible, because the reduced shear correction and its derivatives are relatively small in comparison to the shear spectra and derivatives. In the absence of any corrections, there are near-exact degeneracies which result in large uncertainties when the Fisher matrix is inverted. However, because we are dealing with near-zero eigenvalues in the Fisher matrix, even subtle changes to the models that encode information can significantly change the resulting parameter constraints.

Since the reduced shear correction depends on the observed density of baryonic matter, including it slightly improves the constraint on $\Omega_{\rm b}$. Also, the predicted uncertainties on $h$ are also reduced, as the correction term has an additional factor of the lensing kernel relative to the angular power spectrum; increasing sensitivity to $h$ by a power of two. The fitting formulae used to describe the matter bispectrum, as part of the correction term, also have a non-trivial dependence on $n_{\rm s}$. This means that the sensitivity to $n_{\rm s}$ is also increased, when the correction is made.

\begin{table*}[t]
\centering
\caption{Predicted uncertainties for the $w$CDM parameters from Table \ref{tab:cosmology}, for \emph{Euclid}, in the various cases studied. The `with correction' uncertainties are for the cases when the stated corrections are included Fisher matrix calculation. `RS' denotes reduced shear, and `MB' denotes magnification bias. The combined contribution to the uncertainty from both corrections is labelled `RS+MB'.}
\label{tab:unctab}
\begin{tabular}{c c c c c}
\hline\hline
Cosmological & W/o Correction & With RS Correction  & With MB Correction & With RS+MB Corrections \\
Parameter & Uncertainty (1-$\sigma$) & Uncertainty (1-$\sigma$) & Uncertainty (1-$\sigma$) & Uncertainty (1-$\sigma$) \\
\hline
$\Omega_{\rm m}$& 0.012 & 0.012 & 0.012 & 0.013\\
$\Omega_{\rm b}$& 0.021 & 0.019 & 0.017 & 0.017\\
$h$& 0.13 & 0.092 & 0.081 & 0.082\\
$n_{\rm s}$& 0.032 & 0.019 & 0.018 & 0.018\\
$\sigma_8$& 0.012 & 0.011 & 0.012 & 0.012\\
$\Omega_{\rm DE}$& 0.050 & 0.063 & 0.059 & 0.068\\
$w_0$& 0.17 & 0.15 & 0.14 & 0.17 \\
$w_a$& 0.95 & 0.91 & 0.84 & 1.01 \\
\hline\hline
\end{tabular}
\end{table*}
\begin{table*}[t]
\centering
\caption{Biases induced in the $w$CDM parameters of Table \ref{tab:cosmology}, from neglecting the various corrections, for \emph{Euclid}. The biases when only the basic reduced shear correction is used, when only the basic magnification bias correction is used, when the combined bias from these two corrections is used, and when the IA-enhanced lensing bias correction is used, are given. `RS' denotes reduced shear, and `MB' denotes magnification bias. The combined effect is labelled `RS+MB'.}
\label{tab:nonIAbiastab}
\begin{tabular}{c c c c c}
\hline\hline
Cosmological & Basic RS Correction & Basic MB Correction & Combined RS+MB & IA-enhanced Correction \\
Parameter & Cosmology Bias/$\sigma$ & Cosmology Bias/$\sigma$ & Cosmology Bias/$\sigma$ & Cosmology Bias/$\sigma$ \\
\hline
$\Omega_{\rm m}$ & $-0.11$ & $-$0.43 & $-$0.53 & $-$0.62 \\
$\Omega_{\rm b}$ & 0.016 & $-$0.22 & $-$0.20 & $-$0.25 \\
$h$ & 0.069 & $-$0.029 & 0.040 & $-$0.007 \\
$n_{\rm s}$ & $-$0.093 & $-$0.24 & $-$0.34 & $-$0.27 \\
$\sigma_8$ & 0.068 & 0.36 & 0.43 & 0.52 \\
$\Omega_{\rm DE}$ & 0.31 & 1.05 & 1.36 & 1.32 \\
$w_0$& $-$0.32 & $-$0.35& $-$0.68 & $-$0.67 \\
$w_a$& 0.39 & 0.81 & 1.21 & 1.14 \\
\hline\hline
\end{tabular}
\end{table*}
On the other hand, the uncertainty on $\Omega_{\rm DE}$ worsens upon correcting for the reduced shear approximation. This stems from the fact that the derivative of the correction term with respect to $\Omega_{\rm DE}$ is negative, as a higher dark energy density results in a Universe that has experienced a greater rate of expansion, and accordingly is more sparsely populated with matter. Then, convergence in general is lower, and the magnitude of the correction drops as the approximation is more accurate. Therefore, the magnitude of the reduced shear correction and the strength of the $\Omega_{\rm DE}$ signal are inversely correlated. This means that in the case where the reduced shear correction is made, $\Omega_{\rm DE}$ is less well constrained than in the case where there is no correction. Conversely, increasing $w_0$ and $w_a$ decreases the rate of expansion of the Universe, and so sensitivity to $w_0$ and $w_a$ increases in the case when the correction is made.

\subsection{\label{subsec:MBres}The magnification bias correction}

Figure \ref{fig:Comps_dcl} shows the magnitude of the basic magnification bias correction, relative to the shear auto-correlation spectra (again excluding shot noise). In this case, the relative magnitude of the correction again increases with redshift. However, in the two lowest redshift bins shown, the correction is subtractive. This is the case for the five lowest redshift bins, of the ten that we consider. This is due to the dilution of galaxy density dominating over the magnification of individual galaxies, as there are fewer intrinsically fainter galaxies at lower redshifts. Conversely, at higher redshifts, there are more fainter sources which lie on the threshold of the survey's magnitude cut, that are then magnified to be included in the sample. 

The change in the uncertainty of the cosmological parameters if magnification bias is corrected for, and the bias in these parameters if magnification bias is neglected, are given in Table \ref{tab:unctab} and Table \ref{tab:nonIAbiastab}, respectively.  Accordingly, correcting for the magnification bias has a noticeable effect on the uncertainties of the parameters $\Omega_{\rm b}$, $h$, $n_{\rm s}$, $\Omega_{\rm DE}$, $w_0$, and $w_a$. These changes follow the same trends as those seen from the reduced shear correction. We note, however, that the changes in uncertainty induced by the inclusion of these correction will likely be dwarfed by those resulting from the combination of \emph{Euclid} weak lensing data with other probes; both internal and external. For example, the combination of weak lensing with other \emph{Euclid} probes alone, such as photometric and spectroscopic galaxy clustering as well as the cross-correlation between weak lensing and photometric galaxy clustering, will significantly improve parameter constraints \citep{ISTforecast}.

If magnification bias is not corrected for, the values determined for the parameters $\Omega_{\rm m}$, $\sigma_8$, $\Omega_{\rm DE}$, $w_0$, and $w_a$ are significantly biased at $-0.43\sigma$, $0.36\sigma$, $1.05\sigma$, $-0.35\sigma$, and $0.81\sigma$, respectively. All of these biases are higher than the corresponding bias from making the reduced shear approximation. Given that half of the cosmological parameters are significantly biased if magnification bias is neglected, this correction is necessary for \emph{Euclid}.

\subsection{\label{subsec:combres}The combined correction}

The relative magnitude of the combined reduced shear and magnification bias correction is shown in Fig. \ref{fig:Comps_dcl}. At the lowest redshifts considered, the subtractive magnification bias correction essentially cancels out the reduced shear correction. Then, at intermediate redshifts, the magnification bias is additive and comparable to the reduced shear correction. However, the dominant part of combined corrections is found at the highest redshifts, where the magnification bias correction is particularly strong. Therefore, the combined correction term is predominantly additive across the survey's redshift bins. The effects of the combined corrections, on the predicted cosmological parameter constraints, are stated in Table \ref{tab:unctab} and shown in Fig. \ref{fig:bias_nonIA}. The constraints largely remain affected as they were before. The constraints on $h$ worsen slightly when the two corrections are considered together, due to their differing behaviour at lower redshifts. The uncertainty on $\Omega_{\rm DE}$ also increases further. Additionally, Fig. \ref{fig:bias_nonIA} and Table \ref{tab:nonIAbiastab} show the biases induced in the cosmological parameters if these corrections are neglected. As expected, the biases add together linearly, and when combined the bias on $n_{\rm s}$ also becomes significant. Now, all but two of the cosmological parameters are significantly biased, emphasising the need for these two corrections to be applied to the angular power spectra that will be obtained from \emph{Euclid}.

Furthermore, the combination of weak lensing with other probes will improve parameter constraints, whilst leaving the biases resulting from reduced shear and magnification bias unchanged; meaning that the relative biases in this scenario will be even higher. This further stresses the importance of these corrections.

\subsection{\label{subsec:IAEb}The IA-enhanced lensing bias correction}
\begin{figure}[b]
\centering
\includegraphics[width=1.0\linewidth]{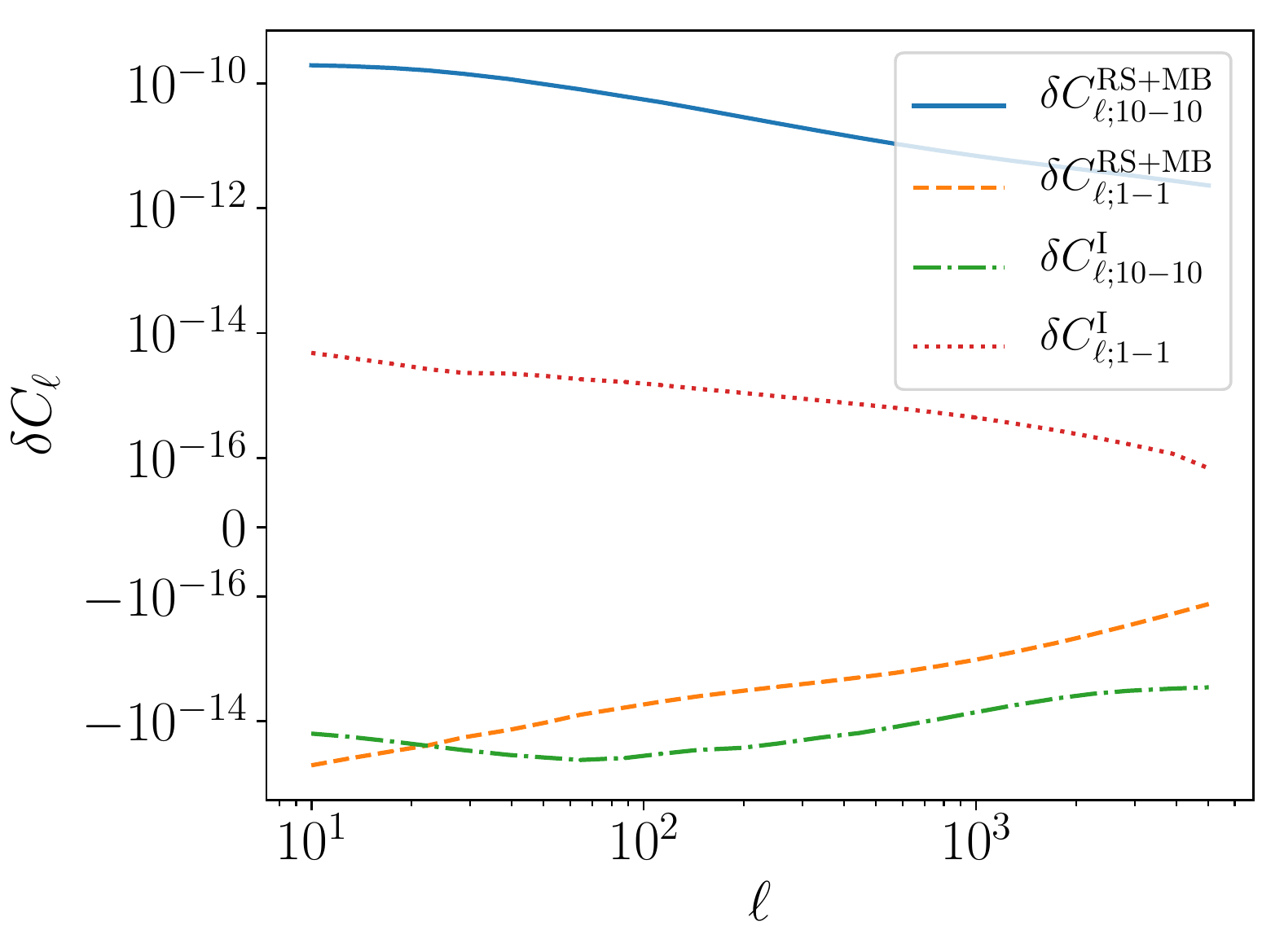}
\caption{Two components of the IA-enhanced lensing bias correction, Eq. (\ref{eq:dCl}) and Eq. (\ref{eq:dClIA}), for the cross-spectra of our first (0.001$\leq z \leq$0.418), and tenth bins (1.576$\leq z \leq$2.50). For the first bin, the basic correction is already sub-dominant, and the additional IA-enhanced terms cancels it out. In the higher redshift bin, the second term is sub-dominant. This trend persists across all bins.}
\label{fig:dcl_comp_IA}
\end{figure}

When the IA-lensing bias interaction term, from Eq. (\ref{eq:dClIA}), is also accounted for, the biases are minimally altered. These are displayed in Table \ref{tab:nonIAbiastab}. From these, we see that the additional term, is non-trivial, but does not induce significant biases in the cosmological parameters obtained at our current level of precision by itself. However, when combined with the basic reduced shear and magnification terms, it leads to the total bias in $\Omega_{\rm b}$ becoming significant, while the total bias in $n_{\rm s}$ is suppressed to now only be on the threshold of significance. The nature of this additional correction, and its relatively minor impact, is explained by Fig. \ref{fig:dcl_comp_IA}.
This charts the change with $\ell$ and redshift, of the two components of the IA-enhanced lensing bias, $\delta C^{\rm RS+MB}_{\ell;ij}$ and $\delta C^{\rm I}_{\ell;ij}$.

From this, we see that for the lowest redshift bins, the two already small terms cancel each other out and at higher redshifts, the latter term is evidently sub-dominant. Accordingly, while upcoming surveys must make the basic reduced shear and magnification bias corrections to extract accurate information, the IA-enhanced correction is not strictly necessary. 

\subsection{\label{subsec:FMres}Forward modelling comparison}

\begin{figure}[t]
\centering
\includegraphics[width=\linewidth]{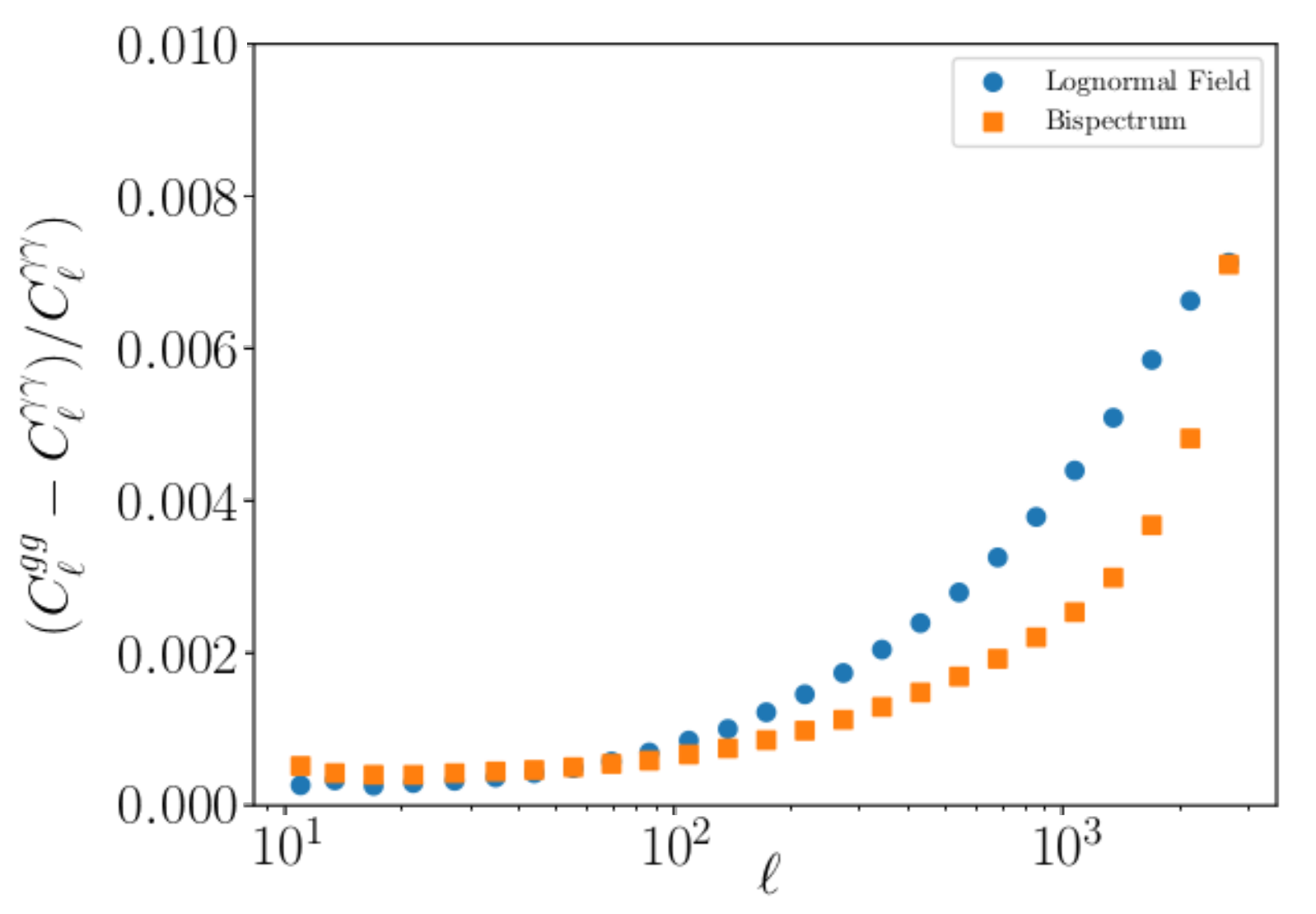}
\caption{Reduced shear correction using the bispectrum perturbative approach (see Sect.~\ref{sec:22}) and using the forward model in the lognormal field approximation presented in~\cite{2019arXiv190405364T}. The data points are plotted at the geometric mean of the $\ell$-bin boundaries. There is mild disagreement at intermediate $\ell$-modes. This is to be expected given the approximations that go into the bispectrum fitting formula and the lognormal field approximation. Nevertheless the agreement at low-$\ell$ and in the highest $\ell$-bin are striking. Here $C_\ell^{gg}$ labels the reduced shear auto-correlation spectrum, while $C_\ell^{\gamma\gamma}$ denotes the shear-shear auto-correlation spectrum.}
\label{fig:compare}
\end{figure}

Figure \ref{fig:compare} compares the reduced shear corrections obtained from the perturbative and forward modelling approaches, for a singular tomographic bin spanning the entire probed redshift range of 0 -- 2.5. There is remarkable agreement between the two within the range of $\ell$-modes that will be observed by \emph{Euclid}, particularly at the highest and lowest $\ell$-modes. We see minor disagreements at intermediate $\ell$-modes, however, this is unsurprising given the various different approximations and assumptions made in the two techniques. We also note that at $\ell$-modes beyond the survey's limit, the lognormal approach will under-predict the perturbative solution. 
Performing cosmological inference on full forward models of the data using density-estimation likelihood-free inference (DELFI)~\citep{alsing2018generalized, alsingndes} to compute the posteriors on cosmological parameters is emerging as a new paradigm in cosmic shear analyses~\citep{2019arXiv190405364T, alsing2018generalized}. It is shown in~\cite{2019arXiv190405364T, alsingndes} that $\mathcal{O}(1000)$ simulations are needed to perform inference on Stage IV data and in contrast to MCMC methods (see e.g.~\cite{foreman2013emcee}) these can be run in parallel, at up to $100$ simulations at a time. In the future it may be easier to handle the reduced shear correction in this paradigm, rather than directly computing the lensing observable with a perturbative expansion. 

The eventual aim for a DELFI pipeline~\citep{2019arXiv190405364T} is to compute lensing observables from full N-body simulations (see e.g~\cite{izard2017ice}). This would avoid the need to write a matter bispectrum emulator trained on simulations, although, the N-body simulations used for this purpose would need to accurately represent the physics of the bispectrum. 

\section{\label{sec:5}Conclusions}

In this work, we quantified the impact that making the reduced shear approximation and neglecting magnification bias will have on angular power spectra of upcoming weak lensing surveys. Specifically, we calculated the biases that would be expected in the cosmological parameters obtained from \emph{Euclid}. By doing so, significant biases were found for $\Omega_{\rm m}$, $\sigma_{8}$, $n_{\rm s}$, $\Omega_{\rm DE}$, $w_0$, and $w_a$ of $-0.53\sigma$, $0.43\sigma$, $-0.34\sigma$, $1.36\sigma$, $-0.68\sigma$, and $1.21\sigma$, respectively. We also built the formalism for an IA-enhanced correction. This was discovered to be sub-dominant. Given the severity of our calculated biases, we conclude that it is necessary to make both the reduced shear and magnification bias corrections for Stage IV experiments.

However, there are important limitations to consider in the approach described here. In calculating these corrections, the Limber approximation is still made. This approximation is typically valid above $\ell\sim100$. But, for \emph{Euclid} we expect to reach $\ell$-modes of ten. Therefore, the impact of this simplification at the correction level must be evaluated. Given that the dominant contributions to the reduced shear and magnification bias corrections come from $\ell$-modes above $100$, we would not expect the Limber approximation to significantly affect the resulting cosmological biases. However, an explicit calculation is still warranted. Furthermore, the various correction terms depend on bispectra which are not well understood: they both involve making a plethora of assumptions, and using fitting formulae that have accuracies of only 30-50$\%$ on small scales. 

In addition, this work does not consider the impact of baryonic feedback on the corrections. We would expect that baryonic feedback behaves in a similar way to lowering the fiducial value $\sigma_8$ (see Appendix \ref{A:VCosmo}), that is, they both suppress structure growth in high density regions. Accordingly, it is likely that the inclusion of baryonic feedback would have an effect on these corrections. If the matter power spectrum is suppressed by a greater fraction than the matter bispectrum, then the biases will increase. However, it is not currently clear to what degree the matter bispectrum is suppressed relative to the matter power spectrum, and existing numerical simulations propose seemingly inconsistent answers (see e.g. \cite{Illustrisbisup} in comparison to \cite{Owlsbisup}). For this reason, we cannot robustly quantify the impact of baryonic feedback on the biases. As knowledge of the impact of baryons on the bispectrum improves, the reduced shear and magnification bias corrections should be modified accordingly. 

An additional hurdle is the large computational expense of these terms; arising from the multiple nested integrals needing numerical computation. Computing the reduced shear and magnification bias corrections for this work took of the order of $24$ hours when multiprocessing across 100 CPU threads. Including the IA-enhanced correction term increases this to $\sim48$ hours. This expense can be prohibitive if the correction is to be included in inference methods. Considering that forward modelling approaches, such as a DELFI pipeline, could both bypass the need for matter bispectrum fitting formulae, and reduce computation time, we recommend that forward modelling should be used to account for these corrections in the future. However, there is also merit in exploring whether the existing processes can be optimised.

\begin{acknowledgements}
We thank the anonymous referee for their constructive comments, and Paniez Paykari for her programming expertise. ACD and TDK are supported by the Royal Society. PLT acknowledges support for this work from a NASA Postdoctoral Program Fellowship and the UK Science and Technologies Facilities Council. Part of the research was carried out at the Jet Propulsion Laboratory, California Institute of Technology, under a contract with the National Aeronautics and Space Administration. The Euclid Consortium acknowledges the European Space Agency and the support of a number of agencies and
  institutes that have supported the development of \emph{Euclid}, in
  particular
the Academy of Finland, the Agenzia Spaziale Italiana,
the Belgian Science Policy, the Canadian Euclid Consortium, the Centre
National d'Etudes Spatiales, the Deutsches Zentrum f\"ur Luft- und
Raumfahrt, the Danish Space Research Institute, the Funda\c{c}\~{a}o
para a Ci\^{e}ncia e a Tecnologia, the Ministerio de Economia y
Competitividad, the National Aeronautics and Space Administration, the
Netherlandse Onderzoekschool Voor Astronomie, the Norwegian Space
Agency, the Romanian Space Agency, the State Secretariat for
Education, Research and Innovation (SERI) at the Swiss Space Office
(SSO), and the United Kingdom Space Agency. A detailed
  complete list is available on the \emph{Euclid}\ web site 
(\url{http://www.euclid-ec.org}).\xspace
\end{acknowledgements}

%
%

\bibliographystyle{aa}
\bibliography{ref}

\begin{appendix}
\section{\label{AGB}Generalised lensing bispectra formulae}

We can extend the methodology used to describe the matter bispectrum, $B_{\delta\delta\delta}$, to describe the bispectrum of three related quantities, $B_{\mu\nu\eta}$. Here, the three fields $\mu$, $\nu$, and $\eta$ are proportional to the density contrast, $\delta$, by some redshift-dependent weightings. This means they behave as $\delta$ would, under a small change in the fiducial cosmology. In this way, the second-order perturbation theory approach of \cite{Fryperturb} remains valid. We also assume Gaussian random initial conditions. Accordingly, the bispectrum is defined by first and second-order terms:
\begin{align}
    \label{eq:bigen0}
    B_{\mu\nu\eta}(\boldsymbol{k}_1, \boldsymbol{k}_2, \boldsymbol{k}_3) &= \langle[\widetilde{\mu}^{(1)}(\boldsymbol{k}_1)+\widetilde{\mu}^{(2)}(\boldsymbol{k}_1)] \nonumber\\
    &\times[\widetilde{\nu}^{(1)}(\boldsymbol{k}_2)+\widetilde{\nu}^{(2)}(\boldsymbol{k}_2)] \nonumber\\
    &\times[\widetilde{\eta}^{(1)}(\boldsymbol{k}_3)+\widetilde{\eta}^{(2)}(\boldsymbol{k}_3)]\rangle,
\end{align}
where the superscripts (2) and (1) denote the second and first-order terms respectively. But because we take Gaussian random initial conditions, the value of the three-point correlation vanishes at the lowest-order. Additionally, we can neglect products of second-order terms, as these are fourth-order terms. Equation (\ref{eq:bigen0}) now becomes: 
\begin{align}
    \label{eq:bigen1}
    B_{\mu\nu\eta}(\boldsymbol{k}_1, \boldsymbol{k}_2, \boldsymbol{k}_3) &= \braket{\widetilde{\mu}^{(2)}(\boldsymbol{k}_1)\widetilde{\nu}^{(1)}(\boldsymbol{k}_2)\widetilde{\eta}^{(1)}(\boldsymbol{k}_3)} \nonumber\\
    &+ \braket{\widetilde{\nu}^{(2)}(\boldsymbol{k}_2)\widetilde{\mu}^{(1)}(\boldsymbol{k}_1)\widetilde{\eta}^{(1)}(\boldsymbol{k}_3)} \nonumber\\
    &+ \braket{\widetilde{\eta}^{(2)}(\boldsymbol{k}_3)\widetilde{\mu}^{(1)}(\boldsymbol{k}_1)\widetilde{\nu}^{(1)}(\boldsymbol{k}_2)}.
\end{align}
The above assumption relating the three fields to $\delta$, also leads us to concluding that $\delta^{(1)}$ is related to $\delta^{(2)}$ in the same way that $\mu^{(1)}$, $\nu^{(1)}$, and $\eta^{(1)}$ are related to $\mu^{(2)}$, $\nu^{(2)}$, and $\eta^{(2)}$ respectively. In which case, we can directly adapt Eq. (40) of \cite{Fryperturb}, to read:
\begin{align}
    \label{eq:bigen2}
    B_{\mu\nu\eta}(\boldsymbol{k}_1, \boldsymbol{k}_2, \boldsymbol{k}_3) &= 2F_2(\boldsymbol{k}_2, \boldsymbol{k}_3)P_{\mu\nu}(\boldsymbol{k}_2)P_{\mu\eta}(\boldsymbol{k}_3) \nonumber\\
    &+ 2F_2(\boldsymbol{k}_1, \boldsymbol{k}_3)P_{\nu\mu}(\boldsymbol{k}_1)P_{\nu\eta}(\boldsymbol{k}_3) \nonumber\\
    &+ 2F_2(\boldsymbol{k}_1, \boldsymbol{k}_2)P_{\eta\mu}(\boldsymbol{k}_1)P_{\eta\nu}(\boldsymbol{k}_2),
\end{align}
with:
\begin{equation}
    \label{eq:Fdef}
    F_2(\boldsymbol{k}_1, \boldsymbol{k}_2) = \frac{5}{7} + \frac{1}{2}\frac{\boldsymbol{k_1}\cdot\boldsymbol{k_2}}{k_1k_2}\bigg(\frac{k_1}{k_2}+\frac{k_2}{k_1}\bigg)+\frac{2}{7}\bigg(\frac{\boldsymbol{k_1}\cdot\boldsymbol{k_2}}{k_1k_2}\bigg)^2.
\end{equation}

As in \cite{Scoccimarro01}, this can then be modified to include numerical fitting to N-body simulations by exchanging $F_2$ for $F_2^{\rm eff}$, as defined in Eq. (\ref{eq:Feff}). The fitting formula determined in \cite{Scoccimarro01} still remains valid, because it does not have any redshift dependence and does not depend on the fiducial cosmology. The density perturbation-IA bispectrum, used in the IA-enhanced lensing bias correction, is then a specific case of this formula, where $\mu=\nu=\delta$, and $\eta=I$.

\section{\label{A:Pess}The `pessimistic' case for \emph{Euclid}}

Given the complexities of modelling both astrophysical uncertainties and the non-Gaussian covariance terms at high $\ell$-modes, \cite{ISTforecast} define a `pessimistic' case for \emph{Euclid} forecasts. In this case, an $\ell$-cut is made at 1500. In this section, we calculate the uncertainties on the cosmological parameters, and the biases induced in them by the reduced shear approximation and magnification when this cut is made. Here, as before, we include both the Gaussian and SSC covariance terms. The results are shown in Table \ref{tab:apppestab}.

Now, we find that the biases are significantly reduced in comparison to the `optimistic' case. However, the bias in $\Omega_{\rm DE}$ remains significant, at $0.28\sigma$. Therefore, even in this non-ideal scenario, the reduced shear and magnification bias corrections must still be made.
\begin{table}[t]
\centering
\caption{Predicted 1-$\sigma$ uncertainties, and biases from neglecting reduced shear and magnification, for the $w$CDM parameters that would be determined from \emph{Euclid}, for the fiducial cosmology of Table \ref{tab:cosmology}, making a scale-cut at $\ell=1500$.}
\label{tab:apppestab}
\begin{tabular}{@{}ccc@{}}
\hline\hline
Cosmological & 1-$\sigma$ & Reduced Shear + Magnification\\
Parameter& Uncertainty & Bias$/\sigma$\\
\hline
$\Omega_{\rm m}$& 0.016 & $-$0.15\\
$\Omega_{\rm b}$& 0.027 & $-$0.073\\
$h$& 0.15 & 0.019\\
$n_{\rm s}$& 0.039 & $-$0.12\\
$\sigma_8$& 0.017 & 0.055\\
$\Omega_{\rm DE}$& 0.13& 0.28\\
$w_0$&  0.24 & $-$0.075\\
$w_a$&  1.82 & 0.22\\
\hline\hline
\end{tabular}
\end{table}

\section{\label{A:VCosmo}The impact of varying the fiducial cosmology}

Owing to the fact that the reduced shear and magnification bias corrections are a projection of the matter bispectrum, while the shear auto and cross-spectra are projections of the matter power spectrum, the relative size of the correction in comparison to the shear spectra is strongly influenced by non-linearity \citep{Shapiro09}. The parameters $\sigma_8$ and $n_{\rm s}$ have the strongest effect on non-linearity, therefore we examine the effect of changing these parameters on the biases, assuming a Gaussian covariance. 
\begin{table}[b]
\centering
\caption{Predicted 1-$\sigma$ uncertainties for the $w$CDM parameters that would be determined from \emph{Euclid} using a Gaussian covariance, for fiducial cosmologies with lower and higher values of $\sigma_8$ and $n_{\rm s}$, (0.716, 0.916) and (0.86, 1.06) respectively, are shown. The uncertainties obtained with the fiducial cosmology of \cite{ISTforecast} (EC19) are shown for reference.}
\label{tab:appunc}
\begin{tabular}{@{}cccccc@{}}
\hline\hline
Cosmo. & EC19 & Low $\sigma_8$ & High $\sigma_8$ & Low $n_{\rm s}$ & High $n_{\rm s}$\\
Param.& 1-$\sigma$ & 1-$\sigma$ & 1-$\sigma$ & 1-$\sigma$ & 1-$\sigma$\\
\hline
$\Omega_{\rm m}$& 0.012 & 0.016 & 0.0085& 0.014 & 0.012\\
$\Omega_{\rm b}$& 0.021 & 0.024 & 0.0043& 0.020 & 0.023\\
$h$& 0.13 & 0.13 & 0.041&  0.12 & 0.13\\
$n_{\rm s}$& 0.031 & 0.031 & 0.012& 0.030 & 0.031\\
$\sigma_8$& 0.011 & 0.014 & 0.041& 0.012 & 0.011\\
$\Omega_{\rm DE}$&  0.050 & 0.065 & 0.037& 0.061 & 0.059\\
$w_0$&  0.16 & 0.21 & 0.13 & 0.17 & 0.16\\
$w_a$&  0.94 & 1.18 & 0.76 & 1.03 & 1.03\\
\hline\hline
\end{tabular}
\end{table}

\begin{figure*}[t]
\centering
\includegraphics[width=\textwidth]{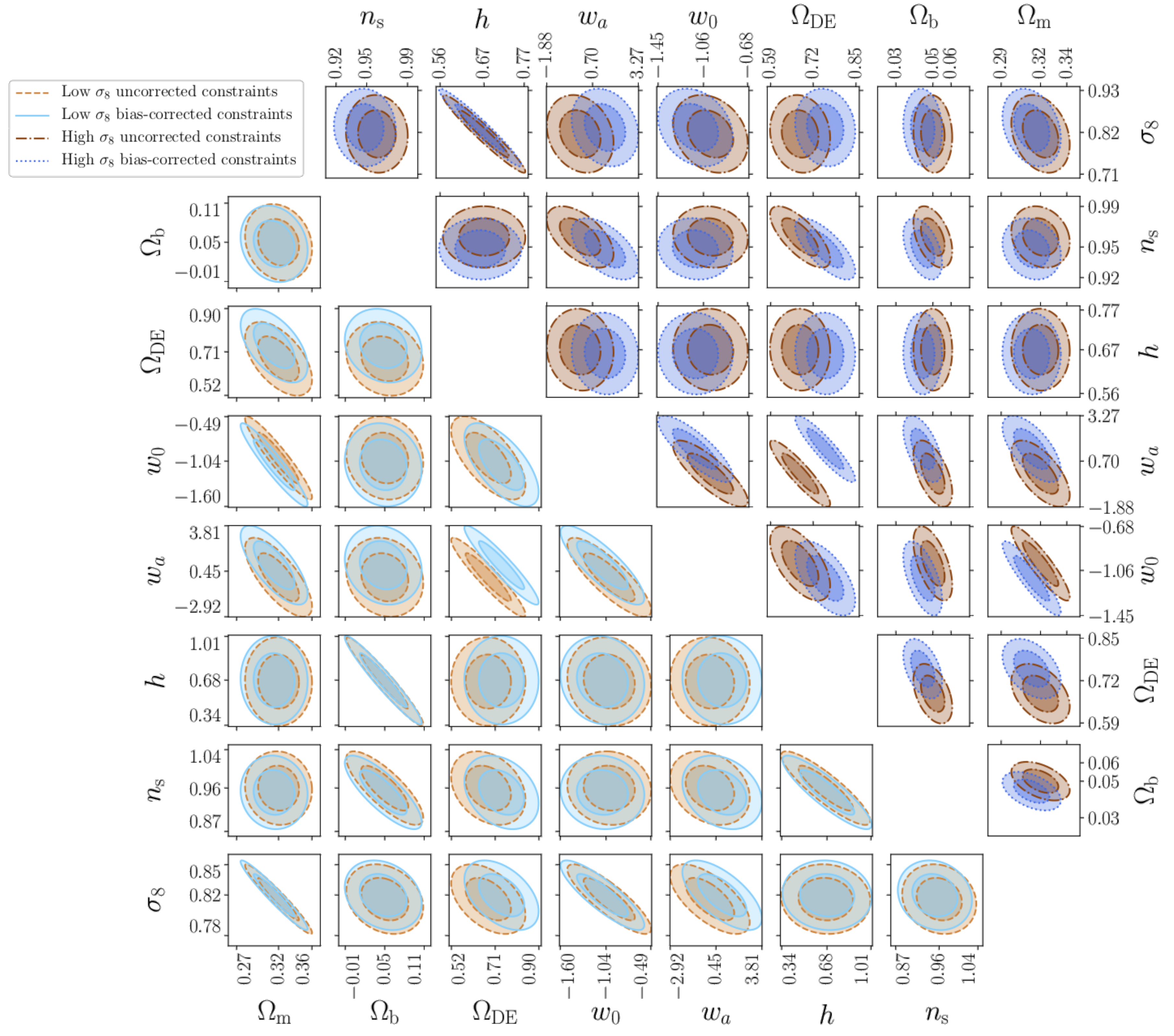}
\caption[width=2.0\linewidth]{Predicted 1- and 2-$\sigma$ contours on the $w$CDM cosmological parameters, for \emph{Euclid}, for different values of $\sigma_8$. All other parameter values are fixed to those shown in Table \ref{tab:cosmology}. Confidence regions and biases for $\sigma_8=0.716$ (bottom left), and $\sigma_8=0.916$ (top right) are shown. The additional IA-enhanced lensing bias is not included in these results. For the lower value of $\sigma_8$, the biases are supressed slightly, but still remain significant. However, for the higher value of $\sigma_8$, the significances of the biases in the cosmological parameters are heightened.}
\label{fig:bias_s8}
\end{figure*}
\begin{table}[b]
\centering
\caption{Biases induced in the $w$CDM parameters, using a Gaussian covariance, from neglecting the two studied corrections, for lower and higher fiducial values of $\sigma_8$ and $n_{\rm s}$, (0.716, 0.916) and (0.86, 1.06) respectively, are shown. The biases obtained with the fiducial cosmology of \cite{ISTforecast} (EC19) are shown for reference.}
\label{tab:appbi}
\begin{tabular}{@{}cccccc@{}}
\hline\hline
Cosmo. & EC19 & Low $\sigma_8$ & High $\sigma_8$ & Low $n_{\rm s}$ & High $n_{\rm s}$ \\
Param. & Bias/$\sigma$ & Bias/$\sigma$ & Bias/$\sigma$ & Bias/$\sigma$ & Bias/$\sigma$\\
\hline
$\Omega_{\rm m}$ &$-$0.51 & $-0.33$ & $-0.76$ & $-0.70$ & $-0.41$\\
$\Omega_{\rm b}$ &$-$0.19 & $-0.097$ & $-1.29$ & $-0.22$ & $-0.23$\\
$h$ &0.059 & 0.076 & $-0.24$ & 0.10 & 0.018\\
$n_{\rm s}$ &$-$0.36 & $-0.29$ & $-0.97$ & $-0.44$ & $-0.50$\\
$\sigma_8$ &0.37 & 0.28 & 0.41 & 0.54 & 0.20\\
$\Omega_{\rm DE}$ &1.36 & 0.89 & 2.07 & 1.31 & 1.43\\
$w_0$ &$-$0.66 & $-0.41$ & $-0.99$ & $-0.67$ & $-0.62$\\
$w_a$ &1.21 & 0.76 & 1.85 & 1.15 & 1.26\\
\hline\hline
\end{tabular}
\end{table}
Table \ref{tab:appunc} and Table \ref{tab:appbi} show the recomputed uncertainties and biases, respectively, when the fiducial values of $\sigma_8$ are lowered to 0.716, and raised to 0.916. These biases are also visualised in Fig. \ref{fig:bias_s8}. As expected, lowering the fiducial value of $\sigma_8$ suppresses the biases, though they still remain significant, whilst raising this value aggravates the biases. Contributing to these changes is also the fact that the predicted uncertainties in the parameters generally decrease as $\sigma_8$ is increased, with the exception of $\sigma_8$ itself. 

\begin{figure*}[p]
\centering
\includegraphics[width=\textwidth]{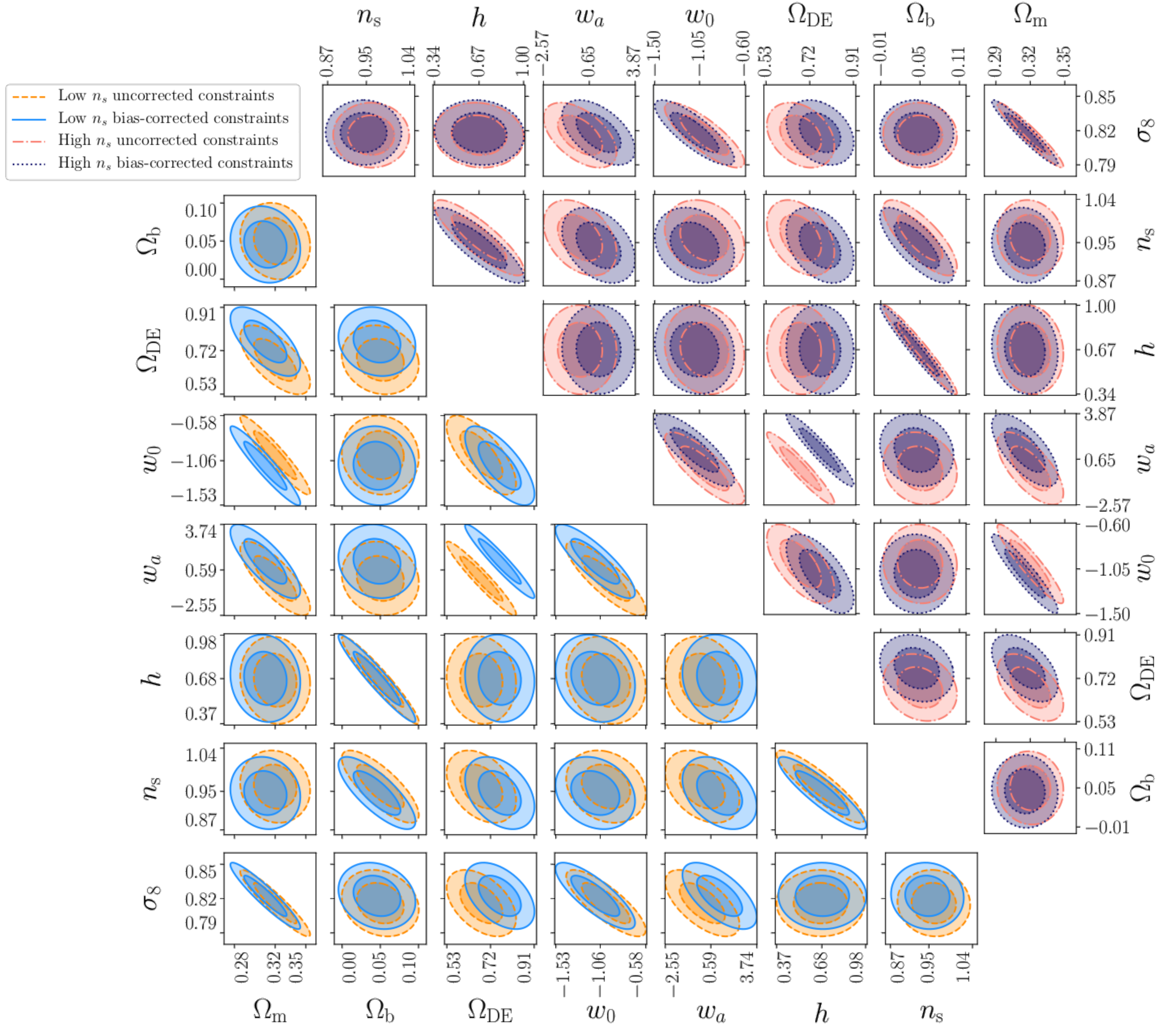}
\caption[width=2.0\linewidth]{Predicted 1- and 2-$\sigma$ contours on the $w$CDM cosmological parameters, for \emph{Euclid}, for different values of $n_{\rm s}$. All other parameter values are fixed to those shown in Table \ref{tab:cosmology}. Confidence regions and biases for $n_{\rm s}=0.86$ (bottom left), and $n_{\rm s}=1.06$ (top right) are shown. The additional IA-enhanced lensing bias is not included in these results. In general, varying the fiducial value of $n_{\rm s}$ does not cause notable change to the significances of the biases.}
\label{fig:bias_ns}
\end{figure*}

The effects on the uncertainties of varying $n_{\rm s}$, to 0.86 then 1.06, are shown in Table \ref{tab:appunc}. Figure \ref{fig:bias_ns} and Table \ref{tab:appbi} show the biases after this variation. The effect on the significances of the biases is less straightforward in this case. The parameters are affected relatively differently in comparison to the variation of $\sigma_8$. In general, the change in the ratio of the biases to the uncertainties is non-trivial, but relatively subtle. The exceptions to this being $\sigma_8$ and $\Omega_{\rm m}$. For these parameters, the biases reduce considerably in magnitude. Despite the changes, the biases in each of the previously affected parameters remain significant.

\end{appendix}

\end{document}